\documentclass[aps,prd,10pt,longbibliography,showkeys,floatfix,showpacs,nofootinbib]{revtex4-1}
\usepackage{amsfonts}%
\usepackage{amsmath}%
\usepackage{amssymb}%
\usepackage{latexsym}%
\usepackage{graphicx}%
\usepackage{hyperref}

\providecommand{\U}[1]{\protect\rule{.1in}{.1in}}

\newcommand{\bmath}{\begin{displaymath}}
\newcommand{\emath}{\end{displaymath}}
\newcommand{\bite}{\begin{itemize}}
\newcommand{\eite}{\end{itemize}}

\renewcommand{\o}{\omega}

\newcommand{\eps}{\varepsilon}

\newcommand{\one}{{1\!\!1}}
\newcommand{\half}{{\textstyle \frac{1}{2}}}

\renewcommand{\o}{\omega}

\newcommand{\bx}{\mathbf{x}}
\newcommand{\by}{\mathbf{y}}

\newcommand{\bk}{\mathbf{k}}
\newcommand{\bq}{\mathbf{q}}

\newcommand{\doubleint}{\int\hspace{-.9em}\int}

\newcommand{\bel}[1]{\begin{equation}\label{#1}}
\newcommand{\bal}[1]{\begin{eqnarray}\label{#1}}
\newcommand{\ee}{\end{equation}}
\newcommand{\ea}{\end{eqnarray}}
\newcommand{\equ}[1]{~Eq.(\ref{#1})}
\newcommand{\vev}[1]{\langle #1\rangle}

\newcommand{\tpsi}{{\tilde\psi}}
\newcommand{\fig}[1]{~Fig.~\ref{#1}}
\renewcommand{\d}[1]{\frac{d#1}{(2\pi)^2}}
\newcommand{\drop}[1]{}

\begin{document}
\title[NCasimir]{A Field Theoretic Approach to Roughness Corrections}
\author{Hua Yao Wu and Martin Schaden}
\affiliation{Department of Physics, Rutgers University, 101 Warren Street, Newark NJ 07102}

\begin{abstract}
We develop a systematic field theoretic description for the roughness correction to the Casimir free energy of parallel plates. Roughness is modeled by specifying a generating functional for correlation functions of the height profile, the two-point correlation function being characterized by the variance, $\sigma^2$, and correlation length, $\ell$, of the profile. We obtain the partition function of a massless scalar quantum field interacting with the height profile of the surface via a $\delta$-function potential.  The partition function of this model also is given by a holographic reduction to three coupled scalar fields on a two-dimensional plane. The original three-dimensional space with a flat parallel plate at a distance $a$ from the rough plate is encoded in the non-local propagators of the surface fields on its boundary. Feynman rules for this equivalent $2+1$-dimensional model are derived and its counter terms constructed. The two-loop contribution to the free energy of this model gives the leading roughness correction. The absolute separation, $a_\text{eff}$, to a rough plate is measured to an equivalent plane that is displaced a distance  $\rho\propto \sigma^2/\ell$ from the mean of its profile. This definition of the separation eliminates corrections to the free energy of order $1/a_\text{eff}^4$ and results in a unitary model. We derive an effective low-energy theory in the limit $\ell\ll a$. It gives the scattering matrix and equivalent planar surface of a very rough plate in terms of the single length scale $\rho$. The Casimir force on a rough plate is found to always \emph{weaken} with decreasing correlation length $\ell$. The two-loop approximation to the free energy interpolates between the free energy of the effective low-energy model and that of the proximity force approximation -- the force on a very rough Dirichlet plate with $\sigma\gtrsim 0.5\ell$ being weaker than on a flat plate at any separation.
\end{abstract}





\startpage{1}
\endpage{120}
\maketitle

\section{Introduction}
Casimir originally\cite{Casimir19481} obtained the force due to electromagnetic zero-point fluctuations between two large ideal parallel metallic flat surfaces at vanishing temperature. His approach was soon generalized to dielectric surfaces\cite{Lifshitz19561,*Lifshitz19562}, finite temperature\cite{Brown19691}, and experimentally more accessible geometries\cite{Derjaguin19581}. The influence of surface roughness was considered only much later\cite{Maradudin19801,*Mazur19811, Novikov19901,*Novikov19902,*Novikov19921,*Novikov19922}, perhaps because this correction was insignificant in early Casimir experiments. When Casimir forces were accurately measured with atomic force microscope techniques\cite{Mohideen19981} at plate separations of only a few hundred nanometers, corrections caused by the roughness of the plates could no longer be ignored. They are even more important at the small separations and higher accuracy of more recent experiments\cite{Decca20031,*Decca20071,Zwol20071,*Zwol20081,*Zwol20082}.  An increasing amount of experimental\cite{Decca20031, *Decca20071,Zwol20071,*Zwol20081,*Zwol20082} and theoretical\cite{Klimchitskaya19991,Emig20031,Genet20031,Neto20051,*Neto20061,Palasantzas20051, Zwol20071,*Zwol20081,*Zwol20082,Bordag2009bk,Broer20111} effort has since been devoted to understanding roughness effects. Currently the only rigorous non-perturbative approach is the proximity force approximation(PFA) (and some recent modifications thereof\cite{Broer20111}). It is accurate when the correlation length $\ell$ of the profile greatly exceeds the separation $a$ of the plates\cite{Klimchitskaya19991,Bordag2009bk}.  Most other approaches consider perturbative corrections to the Green\rq{}s function in powers of $\sigma/a$. The limit of extremely rough plates with $a\gg\ell$ was first obtained using methods of stochastic calculus\cite{Maradudin19801,*Mazur19811}.

For stochastic roughness all perturbative calculations \cite{Novikov19901,Genet20031,*Neto20051,*Neto20061,Zwol20071} show an increase in magnitude of the Casimir energy and force with decreasing correlation length $\ell$, approaching the PFA for $\ell\gg a$\cite{Bordag2009bk}. For a massless scalar field this trend is shown by the dashed curves in fig.~\ref{PertCorrections}. It is qualitatively similar for the electromagnetic case\cite{Genet20031,*Neto20051,*Neto20061,Zwol20071}.  From the point of view of the multiple scattering expansion to the Casimir energy this strengthening of the force is not intuitive and in fact is physically untenable in the limit $a\gg\ell$: the back-scattering, i.e. echo, from a rough plate generally \emph{decreases} with increasing roughness. The reflection coefficient for scattering off a rough surface therefore ought to be reduced in magnitude from that for perfect reflection off a Dirichlet (or ideal metallic) plate. Since the Casimir energy is related to the trace of the Green's function one semi-classically expects a \emph{reduction} in magnitude of the Casimir energy (and force) for a rough plate and not an enhancement.

This qualitative argument becomes rigorous upon using the $GTGT$-formula\cite{Kenneth20061} for the Casimir energy due to a massless scalar field for a stochastically rough but otherwise ideal (Dirichlet) plate and a  perfectly smooth parallel flat plate. When the plate separation $a$ is large compared to the correlation length $\ell$ of the profile one recovers translational invariance and the reflection coefficient becomes diagonal in the transverse momentum. The Casimir energy per unit area, ${\cal E}(a)$, in this case is given by a dimensionally reduced $gtgt$-formula\cite{Cavero20081,*Cavero20082, Bordag2009bk},
\bel{gTgT}
{\cal E}(a)= \int_0^\infty\frac{\kappa^2 d\kappa}{2\pi^2} \ln(1-t_\text{rough}(\kappa)\bar t(\kappa) e^{-2 a\kappa})\ ,
\ee
where the reduced reflection matrices, $t_\text{rough}$ and $\bar t$, for back-scattering off the two parallel plates for a massless scalar field are functions of the wave-number $\kappa$ only. The reduced scattering matrix $\bar t$ for the flat Dirichlet plate is $\bar t=t_D=1$ and unitarity demands that $|t_\text{rough}(\kappa)|\leq 1=t_D$ for reflection off the rough plate. Inspection of\equ{gTgT} then implies that ${\cal E}(a)$ for interaction with a stochastically rough Dirichlet plate should be \emph{reduced} compared to the Casimir energy per unit area for interaction with a flat Dirichlet plate. This argument does not hold in the regime of large correlation length  of the PFA and is rigorous only in the limit $\ell\rightarrow 0$ in which translational invariance is recovered. A problem arises  because the trend of the perturbative analysis is the opposite and gives a Casimir energy that increases in magnitude beyond all bounds for $\ell\rightarrow 0$.

The perturbative analysis  for electromagnetic fields of\cite{Maradudin19801,*Mazur19811,Novikov19901,*Novikov19902,*Novikov19921,*Novikov19922,
Genet20031,Neto20051,*Neto20061} predicts large increases that appear not to be supported by experiment\cite{Decca20031,*Decca20071,Zwol20071,*Zwol20081,*Zwol20082}. Some  experiments\cite{Zwol20071,*Zwol20081,*Zwol20082} with gold coatings described by $\ell\sim 35\text{nm},\sigma\sim 5\text{nm}$ clearly are in the rough regime with $a\gg\ell$ for separations $a>100\text{nm}$.

This investigation was partly motivated by a desire to reconcile the strengthening of the force with increased roughness observed in all perturbative calculations\cite{Maradudin19801,*Mazur19811,Novikov19901,*Novikov19902,*Novikov19921,*Novikov19922,
Genet20031,Neto20051,*Neto20061, Zwol20071,*Zwol20081,*Zwol20082, Bordag2009bk} with the weakening demanded by unitarity and the multiple-scattering formalism. [We will see that the two approaches differ in the definition of the plate separation.] However, the following field theoretic description goes beyond the original  objective of a more rigorous and non-perturbative description of roughness effects. It provides a framework for a consistent loop expansion and includes temperature effects. The loop expansion is uniform in \emph{both} small parameters $\sigma/a$ and $\sigma/\ell$ and the field theory is interesting in itself. It is holographic\cite{Dvali20001,*Maldacena20051} in the sense of being equivalent to a lower-dimensional field theory on the two-dimensional plane that is a boundary of the original space. The existence of a distant plate in an extra dimension is encoded by non-local propagators of the surface fields in the latter model. Roughness corrections to Casimir energies in this sense can be described by a table-top brane theory. 

\section{The Generating Functional of Roughness Correlations}
\label{Roughness}
We consider the standard Casimir configuration of two parallel flat plates at an average separation that is much less than their transverse dimensions\cite{Casimir19481}. A Cartesian coordinate system with $z$-axis normal to the plates is used to describe this system. The profile function $h(\bx)$ associated with a plate at the mean height $\vev{z}=a$ gives the precise position of this surface as a function of the (two) transverse coordinates\footnote{The normal direction is distinguished and bold-faced letters describe two-dimensional \emph{transverse} vectors in the following.} $\bx=(x,y)$,
\begin{align}\label{def-h}
z(\bx)=a+h(\bx) .
\end{align}

We assume the profile of the plate is without enclosures and that $h(\bx)$ is a single-valued function. It nevertheless is often more practical to characterize a rough plate by just a few low-order correlation functions of its profile than by the profile itself.  For sufficiently large plates, a description in terms of (all) correlation functions in fact is exact. The formalism developed below may, in principle, also be applied to plates whose profile is known precisely.

The  $n$-point correlation functions of the profile $h(\bx)$ for a plate of (large) area $A$ are the averages,
\bal{def-D}
D_1&=&\vev{h(\bx_1)}:=A^{-1}\int_A  h(\bx+\bx_1) d\bx\nonumber\\
D_2(\bx_1-\bx_2)&=&\vev{h(\bx_1)h(\bx_2)}:=A^{-1}\int_A  h(\bx+\bx_1)h(\bx+\bx_2) d\bx\nonumber\\
&\vdots&\\
D_n(\bx_1-\bx_2,\dots,\bx_{n-1}-\bx_n)&=&\vev{h(\bx_1)\dots h(\bx_n)}:=A^{-1}\int_A  h(\bx+\bx_1)\dots h(\bx+\bx_n) d\bx\ .\nonumber
\ea
We have here assumed that the plate is large enough for its boundary to be ignored and have used translational invariance to assert that the correlation functions in this case depend only on \emph{differences} in the transverse coordinates\footnote{For exact translational invariance, the finite flat plate should be replaced by a two-dimensional torus of area $A$.}. Isotropy of the profile yields further restrictions;  the $2$-point correlation function $D_2$ in this case depends only on the \emph{distance} between the two points. We assume that the profile and therefore all $n$-point correlation functions of\equ{def-D} can, at least in principle, be measured when the plate is far removed from any other object. The mean position $\vev{z}=a$ of the plate in\equ{def-h} is fixed by requiring that
\bel{novev}
D_1=\vev{h(\bx)}=0.
\ee

It is convenient to collect all correlation functions of\equ{def-D} in a single generating functional $Z_h[\alpha]$,
\bel{Zh}
Z_h[\alpha]=\sum_{n=2}^\infty \frac{1}{n!}\doubleint \alpha(\bx_1)\alpha(\bx_2)\dots\alpha(\bx_n)D_n(\bx_1,\dots,\bx_n) d\bx_1d\bx_2\dots d\bx_n\ ,
\ee
and directly model $Z_h[\alpha]$ instead of individual correlation functions.  With the restriction of\equ{novev}, the simplest model for a rough plate is entirely determined by the $2$-point correlation $D_2$ of the profile. The generating functional for such a (quadratic) Gaussian model is of the form,
\bel{ZhG}
Z^{(2)}_h[\alpha]=\exp \half\{\alpha|D_2|\alpha\}\ ,
\ee
with,
\bel{scalar2d}
\{\alpha|D_2|\alpha\}:=\doubleint \alpha(\bx)D_2(\bx-\by)\alpha(\by) d\by d\bx\ .
\ee
In general,\equ{ZhG} just gives the leading term in a cumulant expansion of $Z_h$. Stochastic roughness is fully described by the covariance of the profile and a Gaussian model by definition is exact in this case. A Gaussian model will also suffice to extract corrections to the free energy to leading order in the roughness profile.  To leading order in the variance $\sigma^2$ even the effect of corrugated profiles $h_\o(\bx)=\sigma\sin(\o x)$ can be described by a Gaussian model. But the four-point correlation in this case is only half of what the Gaussian model predicts,
\begin{align}
\label{periodic}
D_2^\o(\bx-\by)&=\frac{\sigma^2}{2}\cos(\o(x-y))\ \text{  but  },\\ D_4^\o(\bx_1,\bx_2,\bx_3,\bx_4)&=\half (D_2^\o(\bx_1-\bx_2)D_2^\o(\bx_3-\bx_4)+\nonumber\\
&\ \ \ \ +D_2^\o(\bx_1-\bx_3)D_2^\o(\bx_2-\bx_4)+ D_2^\o(\bx_1-\bx_4)D_2^\o(\bx_2-\bx_3))\ .\nonumber
\end{align}
To correctly describe effects due to a periodic profile to order $\sigma^4$ thus requires inclusion of a $4^\text{th}$ order cumulant. Note that correlations of periodic profiles are not positive definite and have no probabilistic interpretation.

The mathematical basis for a field theoretic approach to roughness is that \emph{any} analytic functional $F[h]$ of the profile $h(\bx)$ with translation invariant coefficients can be evaluated using the generating functional $Z_h[\alpha]$. To see this, first consider the evaluation of a monomial in the Taylor expansion of $F[h]$ for small profiles,
\begin{align}
\label{evalf}
\doubleint & d\bx_1d\bx_2\dots d\bx_n F_n(\bx_1-\bx_2,\dots,\bx_{n-1}-\bx_n) h(\bx_1)h(\bx_2)\dots h(\bx_n) =\nonumber\\
&=\frac{1}{A}\int d\bx \doubleint d\bx_1d\bx_2\dots d\bx_n F_n(\bx_1-\bx_2,\dots,\bx_{n-1}-\bx_n) h(\bx+\bx_1)h(\bx+\bx_2)\dots h(\bx+\bx_n) \nonumber\\
&=\doubleint d\bx_1d\bx_2\dots d\bx_n F_n(\bx_1-\bx_2,\dots,\bx_{n-1}-\bx_n) D_n(\bx_1-\bx_2,\dots,\bx_{n-1}-\bx_n)\\
&=\left. \doubleint d\bx_1d\bx_2\dots d\bx_n F_n(\bx_1-\bx_2,\dots,\bx_{n-1}-\bx_n) \frac{\delta}{\delta\alpha(\bx_1)}\frac{\delta}{\delta\alpha(\bx_2)}\dots\frac{\delta}{\delta\alpha(\bx_n)} Z_h[\alpha]\right|_{\alpha=0}\ .\nonumber
\end{align}
The first equality in\equ{evalf} is due to the translational invariance of the coefficient functions $F_n$  [but assumes no regularity of the profile $h(\bx)$ itself]. No further assumptions are required and\equ{evalf} holds for \emph{any} profile on a sufficiently large plate. The third line in\equ{evalf} implies that the result is proportional to the area $A$.  Assuming that all coefficient functions $F_n$ in the Taylor expansion of $F[h]$ are translation invariant and that the expansion converges for the particular profile,\equ{evalf} implies that one may formally evaluate $F[h]$ by the functional derivative,
\bel{evalF}
F[h]=F[\frac{\delta}{\delta \alpha}] Z_h[\alpha]\Big|_{\alpha=0}\ .
\ee
It remains to (approximately) determine the dependence of the partition function of a massless scalar field on the profiles of parallel plates.

\section{Dependence of the Free Energy of a Massless Scalar on the Profiles of Parallel Plates}
This problem has to some extent been addressed in the calculation of lateral Casimir forces for corrugated plates\cite{Golestanian19971,*Golestanian19981,*Emig20011,*Emig20032, Cavero20081, *Cavero20082}, but regular, one dimensional corrugations are very special. Lateral Casimir forces are automatically finite and vanish if one of the two parallel plates is flat. Here we are interested in the influence of profiles on the \emph{normal} Casimir force. The physical interpretation and consistent subtraction of (divergent) contributions due to the curvature of the profile function is one of our main objectives.
 
As in ref.\cite{Bordag19921}, we model the interaction of the $n^\text{th}$ thermal mode\cite{Fried1972bk,*Becher1984bk,*Kapusta1989bk} of a massless scalar quantum field $\phi_n(\bx,z)$ [corresponding to Matsubara frequency $\xi_n=2\pi n T$]  with two semitransparent parallel plates by  $\delta$-function potentials\footnote{We use natural units $\hbar=c=k_B=1$ throughout.},
\bel{interaction0}
V_\text{int}(\bx,z)=\lambda\delta(z-h(\bx)-a)+\bar\lambda\delta(z-\bar h(\bx))\ .
\ee
Here $h,\bar h\ll a/2$ are the profiles of the two surfaces at average positions $\vev{z}=a$ and $\vev{z}=0$, respectively. $\lambda$ and $\bar\lambda>0$ are the corresponding coupling constants of canonical length dimension $a^{-1}$. The limit $\lambda\rightarrow\infty$ (or $\bar\lambda\rightarrow\infty$)  suppresses tunneling through, and enforces Dirichlet boundary conditions on, the corresponding surface. For finite coupling the plate is semitransparent.

Although this scalar model appears far removed from reality, it is sufficiently simple to analyze thoroughly and does exhibits some features encountered in the electrodynamic case. It in particular essentially describes the thin plate limit of the electric contribution to the Casimir force\cite{Parashar20111,*Parashar20112}.  Since we are not primarily interested in lateral forces\cite{Golestanian19971,*Golestanian19981,*Emig20011,*Emig20032, Cavero20081, *Cavero20082}, we consider the case where only one of the plates is rough by setting $\bar h=0$. This restriction does not qualitatively affect the approach but greatly simplifies the model. Expanding the interaction of\equ{interaction0} for $h(\bx)\ll a$, one arrives at the interaction Hamiltonian,
\begin{subequations}
\label{interaction}
\begin{align}
H_\text{int}[h,\phi]&=H^{(\eps)}[h]+\sum_n\int d\bx \half[\lambda\phi_n^2(\bx,a+h(\bx))+\bar \lambda\phi_n^2(\bx,0)]\label{Hint}\\
&\sim H^{(\eps)}[h]+\sum_n H_\text{int}^{(0)}[\phi_n]+ H_\text{int}^{(1)}[h,\phi_n]+ H_\text{int}^{(2)}[h,\phi_n]+\dots \text{ with }\nonumber\\
H_\text{int}^{(0)}[\phi]&=\int d\bx \left[\frac{\lambda}{2}\phi^2(\bx,a)+\frac{\bar\lambda}{2}\phi^2(\bx,0)\right]\ ,\label{H0}\\
H_\text{int}^{(m)}[h,\phi]&=\frac{\lambda}{2} \int d\bx\, \frac{h^m(\bx)}{m!} \frac{\partial^m}{\partial a^m}\phi^2(\bx,a)\ \ \text{for}\ m>0\ ,\label{Hm}\\
H^{(\eps)}[h]&=\int d\bx\, h(\bx) c^{(\eps)}_1(a;\lambda,\bar\lambda,T)+\frac{1}{2} \doubleint d\bx d\by \,h(\bx) c^{(\eps)}_2(\bx-\by;\lambda)h(\by)+\dots.\label{Hcounter}
\end{align}
\end{subequations}
Note that $H_\text{int}^{(0)}$ describes the interaction of the scalar field with two flat plates and does not depend on the profile $h(\bx)$.  $H_\text{int}^{(m)}$ is of $m^\text{th}$ order in the profile. In $H^{(\eps)}$ we include counter terms of all orders in the profile $h$ that depend on the regularization parameter $\eps$ but not on the quantum field $\phi$. The constant $c^{(\eps)}_1(a;\lambda,\bar\lambda,T)$ of the one-point counter term depends on the plate separation $a$, temperature $T$ and \emph{both} coupling constants $\lambda,\bar\lambda$. This finite counter term enforces the constraint of\equ{novev} at any temperature and separation when the interaction with the scalar $\phi$ is turned on. It ensures that the parameter $a$ represents the \emph{mean} separation of the plates even when $\lambda$ and $\bar\lambda$ do not vanish. [We shall later see that a different definition of the separation is to be preferred.] The coefficient function $c^{(\eps)}_2(\bx-\by;\lambda)$ of the two-point counter term guarantees that the (measured) correlation $\vev{h(\bx)h(\by)}$  at temperature $T=0$ remains $D_2(\bx-\by)$ when the two plates are far apart and $\lambda>0$.  $c^{(\eps)}_2(\bx-\by;\lambda)$ by construction does not depend on the separation $a$, temperature $T$ or the coupling strength $\bar\lambda$ of the distant plate. The $(n>1)$-point counter terms ensure that the corresponding $n$-point correlation of the profile also remains unchanged at $T=0$ when the plates are far apart and the interaction with the scalar is switched on. These counter terms do not depend on $a,T, or \bar\lambda$ and are oblivious to the existence of another plate, but diverge in the limit  $\eps\rightarrow 0+$ in which the regularization is removed. The model requires an infinite number of counter terms and is not renormalizable in the sense of Dyson.  However, it is renormalizable in a more modern sense\cite{Gomis19961}. Determination of the counterterms in fact does  not diminish the predictive power of this model, since we assumed from the outset  that correlation functions of the profile $h$ at $T=0$ are all known (measured) when the other plate is far removed. No counter terms for the quantum field are required and the model unambiguously predicts finite effects due to interaction with another plate for \emph{any given} profile $h(\bx)$ [and its associated correlations] measured at $T=0$.

Thermal correlation functions of a free massless scalar field in equilibrium at temperature $T$ in  Matsubara's formalism\cite{Fried1972bk,*Becher1984bk,*Kapusta1989bk} are generated by
\bel{Z0phi}
Z_0[j;T]=\exp\left[-\frac{1}{T}{\cal F}^{(0)}+\frac{T}{2}\sum_n (j_n|G^0_n|j_n)\right]\ ,
\ee
where  ${\cal F}^{(0)}=-\frac{\pi^2 T^4 V}{90}$ is the Helmholtz free energy of a massless scalar field in a 3-dimensional Euclidean space of volume $V$ and\footnote{The scalar product $\{\dots\}$ defined in\equ{scalar2d} differs in its function space from  $(\dots)$ given in\equ{scalarprod}. We notationally differentiate between three-dimensional vectors $\vec{u}$ and two-dimensional vectors ${\bf u}$.}
\bel{scalarprod}
(j_n|G^0_n|j_n):=\int d^3x  \int d^3y  j_n(\vec{x}) G^0_n(\vec{x}-\vec{y}) j_n(\vec{y})\ .
\ee
The free thermal Greens-function,
\bel{G0}
G^0_n(\vec{x}-\vec{y}) = \frac{e^{-2\pi n T|\vec{x}-\vec{y}|}}{4\pi|\vec{x}-\vec{y}|} ,
\ee
satisfies the differential equation,
\bel{G0e}
(\xi_n^2-\nabla^2) G^0_n(\vec{x}-\vec{y})=\delta(\vec{x}-\vec{y})\ ,\text{    with  } \xi_n=2\pi n T.
\ee

The generating function of thermal Greens-functions at temperature $T$ of the interacting model is\cite{Fried1972bk,*Becher1984bk,*Kapusta1989bk},
\bal{GenZ}
Z[j,h;T,a]&=&\exp\left[-\frac{1}{T} H_\text{int}[h,\frac{\delta}{\delta j}]\right]\; Z_0[j;T]\\
&\sim & \exp \frac{-1}{T}\left[H^{(\eps)}[h]+\sum_{m=1} \sum_n H_\text{int}^{(m)}[h,\frac{\delta}{\delta j_n}]\right] Z^\parallel[j;T,a]\ .\nonumber\\
\ea
Here $Z^\parallel[j;T,a]$ generates the thermal Green's functions of the scalar field in the presence of two flat parallel plates separated by a distance $a$,
\bal{Za}
Z^\parallel[j;T,a]&=&\exp\left[-\frac{1}{T} \sum_n H_\text{int}^{(0)}[\frac{\delta}{\delta j_n}]\right] Z_0[j;T]\\
&=& \exp\left[-\frac{1}{2T}\sum_n\int d\bx \big[\lambda\frac{\delta^2}{\delta j_n(\bx,a)^2}+\bar\lambda\frac{\delta^2}{\delta j_n(\bx,0)^2}\big]\right]\ Z_0[j;T]\nonumber\\
&=&\exp\left[-\frac{1}{T}{\cal F}^\parallel(T;a,\lambda,\bar\lambda)+\frac{T}{2}\sum_n (j_n|G^\parallel_n|j_n)\right]\ .\nonumber
\ea
The free-energy ${\cal F}^\parallel$ of a massless scalar field in the presence of two semi-transparent parallel plates was obtained in\cite{Cavero20081,*Cavero20082} and is reproduced in App.~\ref{AppA}. The thermal Green's function, $G_n^\parallel$, of a scalar thermal mode in the presence of two flat parallel plates satisfies the partial differential equation,
\bel{GaTdiffeq}
(\xi_n^2-\nabla^2+\lambda\delta(z-a)+\bar\lambda\delta(z)) G^\parallel_n(\bx-\by,z,z')=\delta(z-z')\delta(\bx-\by)\ ,\text{    with  } \xi_n=2\pi n T.
\ee
Exploiting transverse translational symmetry, $G_n^\parallel$ is expressed by the  dimensionally reduced Green's function $g^\parallel$ as,
\bel{G2def}
\vev{\phi_n(\bx,z)\phi_n(\by,z')}^\parallel=G_n^\parallel(\bx-\by,z,z')=\int \frac{d\bk}{(2\pi)^2}\;e^{i\bk(\bx-\by)} g^\parallel(z,z';\kappa_n),
\ee
with  $\kappa^2_n=\xi^2_n+\bk^2=(2\pi n T)^2+\bk^2$. Inserting\equ{G2def} in\equ{GaTdiffeq} gives the ordinary second order differential equation satisfied by $g^\parallel(z,z';\kappa)$,
\bel{gdiff}
(\kappa^2-\frac{d^2}{dz^2}+\lambda\delta(z-a)+\bar\lambda\delta(z)) g^\parallel(z,z';\kappa)=\delta(z-z')\ .
\ee
The solution to\equ{gdiff} for physical boundary conditions is well known\cite{Cavero20081,*Cavero20082,Shajesh20111} and reproduced in\equ{g2plates} of App.~\ref{AppB}. 

Note that the reduced Green's function $g^\parallel$ for the $n^\text{th}$ Matsubara mode is a function of $\kappa_n$ only. An exponential cutoff,
\bel{cutoff}
g^\parallel(z,z';\kappa)\rightarrow g^{\parallel(\eps)}(z,z';\kappa)= e^{-\eps\kappa} g^\parallel(z,z';\kappa)\ ,
\ee
thus simultaneously regularizes loop integrals and summations. This regularization manifestly preserves transverse translational invariance of the corresponding regularized generating functional $Z^{(\eps)}[j,h;T,a]$. Our renormalization condition that corrections to the correlation functions of the profile vanish at $T=0$  for $a\rightarrow \infty$ also is invariant under transverse translations. This implies that the regularized partition functional $Z^{(\eps)}[0,h;T,a]$ is analytic in the profile $h$ with coefficient functions that  are invariant under transverse translations. We therefore can use\equ{evalF} to evaluate this functional of the profile and express the free energy as,
\begin{align}\label{FreeEnergy}
&&&{\cal F}(T)={\cal F}^\parallel(T)-T\lim_{\eps\to 0^+} \ln {\cal Z}^{(\eps)}[0,0;T,a],\\
\text{where }&&&{\cal Z}^{(\eps)}[0,0;T,a]=\left. Z^{(\eps)}[0,{\textstyle\frac{\delta}{\delta \alpha}};T,a] Z_h[\alpha]\right|_{\alpha=0}\ .\nonumber
\end{align} 
With a Gaussian model describing the roughness correlations, the (regularized) generating  function ${\cal Z}^{(\eps)}$ in\equ{FreeEnergy} is,
\begin{align}\label{PM}
{\cal Z}^{(\eps)}[j,\alpha;T,a]&=\\ 
&\hspace{-4em}=\exp\big[{\textstyle-\frac{1}{T}}  H^{(\eps)}[{\textstyle\frac{\delta}{\delta \alpha}}]{\textstyle-\frac{1}{T}}\sum_{m=1} \sum_n H_\text{int}^{(m)}[{\textstyle \frac{\delta}{\delta \alpha},\frac{\delta}{\delta j}}])\big] \exp \big[{\textstyle {\frac{1}{2}}\{\alpha|D_2|\alpha\}+{\textstyle\frac{T}{2}}\sum_n (j_n|G^{\parallel(\eps)}_n|j_n)}\big]\ .\nonumber
\end{align}
The functional ${\cal Z}^{(\eps)}[\alpha,j;T,a]$ generates the Green's functions of two interacting scalar fields $\phi$ and $h$ that are supported on  Euclidean spaces of different dimension. The base-manifold of the Euclidean quantum field $\phi$ is $S_1\times \mathcal{R}_3$, whereas the roughness field $h$ is only supported on a two-dimensional plane in $\mathcal{R}_3$. From a contemporary point of view this is a miniature brane-world in which a matter field $h(\bx)$ is confined to the two-dimensional universe of a plane embedded in a four-dimensional Euclidean bulk-space supporting $\phi$. Because interactions occur only on the two-dimensional brane, the model is reducible to an equivalent $2+1$-dimensional field theory on the \lq{}boundary\rq{} universe of the plane. Roughness effects first manifest themselves in two-loop contributions to the free energy of this dimensionally reduced model.

\subsection{Feynman rules}
\label{rules}
\equ{PM} defines a perturbative expansion and the associated Feynman rules for the scattering matrix on the  plane. It will be advantageous to derive these rules in transverse momentum space.  The presence of a second (flat) plate leads to non-local parts of propagators that are exponentially suppressed for momenta $\kappa a\gg 1$. They describe the back-scattering off the distant (flat) plate and inform of its presence.

\subsubsection{Propagators}
The model on the plane has four propagators. In transverse momentum space they are given by Eqs.~(\ref{valGa}),(\ref{valpGa}), (\ref{valpGpa}) of appendix~\ref{AppB} and by the Fourier transform $d(\bk)$ of $D_2$. On the two-dimensional plane, $\phi_n(\bx,a)$ and $\frac{\partial}{\partial a}\phi_n(\bx,a)$ are independent and distinct modes.  Introducing their Fourier components,
\bel{Fourierfields}
\psi_n(\bk):=\int d\bx \,e^{i\bk\bx}\phi_n(\bx,a)\ \ \text{and   }\ \ \tpsi_n(\bk):=\int d\bx\, e^{i\bk\bx}\frac{\partial}{\partial a}\phi_n(\bx,a)\ ,
\ee
the four non-vanishing propagators of the surface model in (two-dimensional) Fourier space are,
\begin{subequations}
\label{props}
\begin{align}
\vev{\psi_n(\bk)\psi_n(-\bk)}^\parallel&=g^{(f)}_{00}(\kappa_n)+g^{(s)}_{00}(\kappa_n)&=&\frac{1}{\lambda+2\kappa_n} &-&\frac{2\kappa_n t_n^2\bar t_n e^{-2\kappa_n a} }{\lambda^2\Delta_n},\label{g00}\\
\vev{\psi_n(\bk)\tpsi_n(-\bk)}^\parallel&=g^{(s)}_{01}(\kappa_n)=g^{(s)}_{10}(\kappa_n)&=&&&\frac{\kappa_n t_n \bar t_n e^{-2\kappa_n a}}{\lambda\Delta_n},\label{g01}\\
\vev{\tpsi_n(\bk)\tpsi_n(-\bk)}^\parallel&=g^{(f)}_{11}(\kappa_n)+g^{(s)}_{11}(\kappa_n)&=&-\frac{\kappa_n}{2}&-&\frac{\kappa_n\bar t_n e^{-2\kappa_n a}}{2\Delta_n},\label{g11}\\
\vev{h(\bk)h(-\bk)}&= d^{\,(f)}(\kappa_0)=\int d\bx D_2(\bx) e^{i\bk\bx}&=&2\pi\sigma^2\ell^2 e^{-\ell^2\bk^2/2} ,&&\label{d}
\end{align}
\end{subequations}
with
\bel{definitions}
\Delta_n:=1-t_n\bar t_n e^{-2 \kappa_n a}\ , \ \ t_n:=\frac{\lambda}{\lambda+2\kappa_n}\ , \ \ \bar t_n:=\frac{\bar\lambda}{\bar\lambda+2\kappa_n}\ , \ \ \kappa_n:=\sqrt{(2\pi n T)^2+\bk^2}\ .
\ee
In \equ{props} we have decomposed the propagators into separation-dependent, $(s)$oft parts that are exponentially suppressed for $a\kappa\gg 1$ and $(f)$ast parts that remain for infinite separation $a\rightarrow\infty$. [In the following this distinction is dropped when irrelevant]. Note that  $g_{00}$ and $g_{01}$ vanish in the strong coupling (Dirichlet) limit $\lambda\rightarrow \infty$, whereas correlations of the normal derivative on the surface described by $g_{11}$ do not. There are no transitions between thermal modes in this model and the quantities  $t_n,\bar t_n$ and $\Delta_n$ defined in\equ{definitions} are diagonal and functions of $\kappa_n$  only.  Lacking an experimental determination, the two-point correlation function for the profile is assumed to be given by a normal distribution in\equ{d}. It is characterized by its variance $\sigma^2$ and correlation length $\ell$. Although we will give results only for this particular form of the two-point correlation function, other correlation functions\cite{Palasantzas20051} that vanish faster than any power of the transverse momentum are equally admissible and do not qualitatively alter our considerations and conclusions.

The dependence of the propagators on the exponential cutoff introduced in\equ{cutoff} has been suppressed in \equ{props} but should be implicitly assumed. The cutoff length $\eps$ can be neglected compared to the separation $a$ in all exponentially suppressed $(s)$oft terms and loop integrals containing such a propagator remain finite in the limit $\eps\rightarrow 0^+$. However, $a$-independent parts of internal loops containing only $(f)$ast propagators do diverge and a regularization is formally necessary to evaluate counter terms.

We collect the $\psi,\tpsi$ propagators of\equ{props} in the matrix,
\bel{PropMat0}
{\bf g}(\kappa)=\left(
               \begin{array}{cc}
                 g_{00}(\kappa) & g_{01}(\kappa) \\
                 g_{10}(\kappa) & g_{11}(\kappa) \\
               \end{array}
             \right)\ ,
\ee
whose negative inverse ${\bf\Gamma}^{(0)}$ is the matrix of  two-point vertices for vanishing profile $h=0$,
\bal{GammaMat0}
{\bf\Gamma}^{(0)}(\kappa):=-{\bf g}^{-1}(\kappa)&=&-\left(
               \begin{array}{cc}
                 \lambda+2\kappa(1+\bar t e^{-2\kappa a}) & 2\bar t e^{-2\kappa a} \\
                2\bar t e^{-2\kappa a} & -\frac{2}{\kappa}(1-\bar t e^{-2\kappa a}) \\
               \end{array}
             \right)\\&=&2\left(\begin{array}{cc}
                 -\kappa & 0\\
                0 & \kappa^{-1}\\
               \end{array}
             \right)-\left(\begin{array}{cc}
                 \lambda & 0 \\
                0 & 0\\
               \end{array}
             \right)-2\bar t e^{-2\kappa a}\left(\begin{array}{cc}
                 \kappa & 1 \\
                1 & \kappa^{-1} \\
               \end{array}
             \right)\ .\nonumber\\
\ea
The corresponding generating functional of tree-level two-point vertices is,
\bel{eff0}
\Gamma^{(0)}[\psi,\tpsi]=\frac{1}{2}\sum_n\int \d{\bk} (\psi_n(\bk),\tpsi_n(\bk))\cdot{\bf \Gamma}^{(0)}(\kappa_n)\cdot\left(
               \begin{array}{c}
                 \psi_n(-\bk)\\
               \tpsi_n(-\bk) \\
               \end{array}
             \right)\ ,
\ee
The dependence of ${\bf \Gamma}^{(0)}$ on the coupling $\lambda$ is only in the  $\psi\psi$-component and is linear. A finite effective action implies vanishing $\psi$ but does not constrain $\tpsi$ in the strong coupling (Dirichlet) limit.  Note that the quadratic form $\Gamma^{(0)}[\psi,\tpsi]$ is an indefinite metric on the function space.

The vertex function ${\bf \Gamma}^{(0)}(\kappa_n)$ is diagonal in the Fourier-space of $(\bk,\xi_n)$-modes and
\bel{detG0}
\half\ln [-\det{\bf \Gamma}^{(0)}(\kappa_n)]=\half\ln(\frac{4\Delta_n}{1-t_n})=\half\ln(\Delta_n)+\half\ln(1+\frac{\lambda}{2\kappa_n})+\ln 2\ .
\ee
Comparing with Eqs.~(\ref{freeplate0})~and~(\ref{parallel0}) shows that  \equ{detG0} essentially is the contribution to the free energy of a thermal mode in the presence of two parallel flat plates. \equ{detG0} includes the contribution to the free energy due  to the plate itself but not that of the other (distant) plate. This correspondence is further evidence that the (negative) effective action for the surface modes of a flat plate is indeed given by\equ{eff0}.

\subsubsection{Vertices}
\label{vertices}
Because the interaction in\equ{interaction} is quadratic in the scalar $\phi$ and the profile $h(\bx)$ does not depend on time, primitive vertices are diagonal in the Matsubara frequency and we need only specify their dependence on $\kappa_n=\sqrt{\xi_n^2+\bk^2}$ and $\kappa_n^\prime=\sqrt{\xi_n^2+\bk^{\prime\, 2}}$.

\begin{figure}
\includegraphics[scale=0.5]{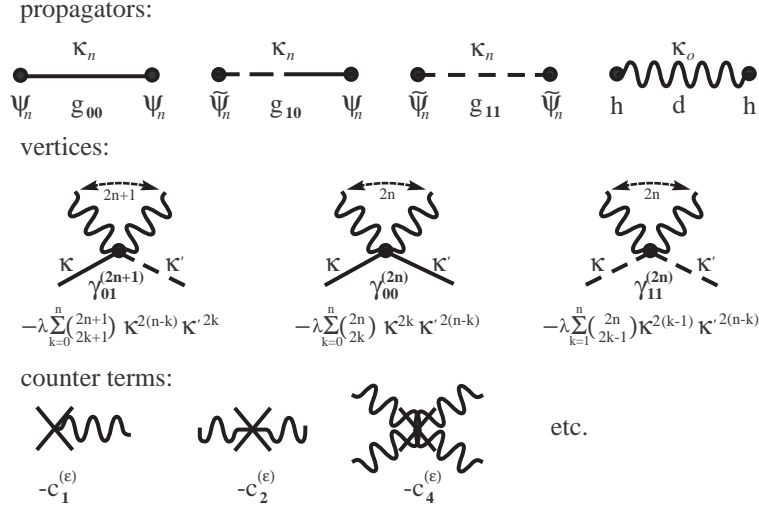}
\caption{Propagators, vertices and counter terms of the $2+1$ dimensional field theory on the planar surface. The \lq{}roughness field\rq{} $h$ corresponds to wavy- and the two dynamical surface fields to solid- and dashed- lines. Counter term vertices are depicted as crosses. Apart from $c_1$, the theory only requires counter terms with an \emph{even} number of external $h$-legs. See the main text for details.}
\label{FeynmanGraphs}
\end{figure}

The interaction $H_\text{int}^{(1)}$ in\equ{Hm} leads to transitions between the $\psi$ and $\tpsi$-modes. It corresponds to the three-point vertex $\gamma^{(1)}_{10}$ in fig.~\ref{FeynmanGraphs},
\bel{V3}
\gamma^{(1)}_{01}(\kappa,\kappa^\prime)=\gamma^{(1)}_{10}(\kappa^\prime,\kappa)=-\lambda\ .
\ee
Expressions for primitive $(m+2)$-point vertices $\gamma^{(m)}$ with $m$ external roughness profiles are similarly obtained from\equ{Hm} by noting that in Fourier space $(\partial/\partial a)^m\phi_n(\bk,a)$ may be replaced by $\kappa_n^2 (\partial/\partial a)^{m-2}\phi_n(\bk,a)$ due to\equ{gdiff}. Primitive vertices with an odd number of profiles lead to transitions between $\psi$ and $\tpsi$ modes and are given by,
\bel{oddv}
\gamma^{(2n+1)}_{01}(\kappa,\kappa^\prime)=\gamma^{(2n+1)}_{10}(\kappa^\prime,\kappa)=-\lambda\sum^{n}_{k=0}\binom{2n+1}{2k}\kappa^{2 k}\kappa^{\prime\, 2(n-k)}\ .
\ee
Vertices with an even number of profiles do not cause transitions between $\psi$ and $\tpsi$ fields and come in two kinds,
\bel{evenv}
\gamma^{(2n)}_{00}(\kappa,\kappa^\prime)=-\lambda\sum^{n}_{k=0}\binom{2n}{2k}\kappa^{2k}\kappa^{\prime\, 2(n-k)}\ \ \text{and}\ \  \gamma^{(2n)}_{11}(\kappa,\kappa^\prime)=-\lambda\sum^{n}_{k=1}\binom{2n}{2k-1}\kappa^{2(k-1)}\kappa^{\prime\, 2(n-k)} \ .
\ee
In diagrammatic form these vertices are shown in fig.~\ref{FeynmanGraphs}.

Introducing the Fourier-transform $h(\bk)=\int d\bx e^{i\bk\bx}h(\bx)$ of the profile, the primitive vertices may be collected to vertex functionals generating the interactions of the $n^\text{th}$ Matsubara mode with the profile,
\bal{Vertex}
\lambda V_n^{00}(\bk,\bk^\prime)&:=&\sum_{m=1}^\infty\frac{(2\pi)^2}{(2m)!}\left[\prod_{j=1}^{2m}\int\frac{d\bk_j}{(2\pi)^2} h(\bk_j)\right]\delta(\bk+\bk^\prime+\sum_{j=1}^{2m}\bk_j)\; \gamma^{(2m)}_{00}(\kappa_n,\kappa_n^\prime)\nonumber\\
\lambda V_n^{01}(\bk,\bk^\prime)&:=&\sum_{m=0}^\infty\frac{(2\pi)^2}{(2m+1)!}\left[\prod_{j=1}^{2m+1}\int\frac{d\bk_j}{(2\pi)^2} h(\bk_j)\right]\delta(\bk+\bk^\prime+\sum_{j=1}^{2m+1}\bk_j)\; \gamma^{(2m+1)}_{01}(\kappa_n,\kappa_n^\prime)\nonumber\\
V_n^{10}(\bk,\bk^\prime)&:=&V_n^{01}(\bk^\prime,\bk)\\
\lambda V_n^{11}(\bk,\bk^\prime)&:=&\sum_{m=1}^\infty\frac{(2\pi)^2}{(2m)!}\left[\prod_{j=1}^{2m}\int\frac{d\bk_j}{(2\pi)^2} h(\bk_j)\right]\delta(\bk+\bk^\prime+\sum_{j=1}^{2m}\bk_j)\; \gamma^{(2m)}_{11}(\kappa_n,\kappa_n^\prime)\ .\nonumber
\ea
Together with\equ{GammaMat0} the interactions of\equ{Vertex} determine the vertex functional  $\Gamma[\psi,\tpsi;h]$ for any given profile $h(\bx)$,
\begin{align}
\label{Gammah}
\Gamma[\psi,\tpsi;h]&=\Gamma^{(0)}[\psi,\tpsi] +\frac{\lambda}{2} \sum_n \int \frac{d\bk d\bk^\prime}{(2\pi)^4}(\psi_n(\bk),\tpsi_n(\bk))\cdot{\bf V}_n[h](\bk,\bk^\prime) \cdot\left(
               \begin{array}{c}
                 \psi_n(\bk^\prime)\\
               \tpsi_n(\bk^\prime) \\
               \end{array}
             \right)\nonumber\\
\text{where}\ \ & {\bf V}_n[h]=\left(\begin{array}{cc}
                 V_n^{00} & V_n^{01}\\
               V_n^{10}& V_n^{11}\\
               \end{array}
             \right)\ .
\end{align}

\subsubsection{Counter terms}
\begin{figure}
\includegraphics[scale=0.70]{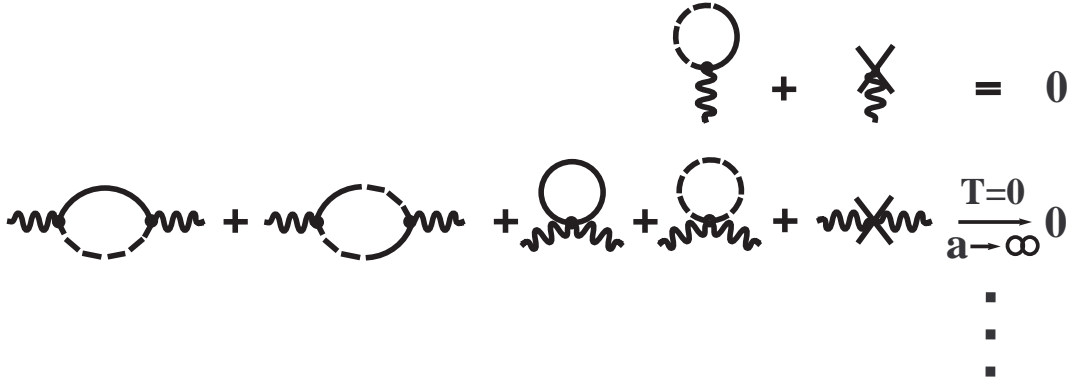}
\caption{{\small Feynman graphs for $c_1$ and $c_2$ counter-terms to one loop. $c_1$ is finite but eliminates all tadpole contributions and guarantees that $\vev{h}=0$ for any coupling, temperature and separation. Counter terms $c_2, c_4,\dots$ are local and guarantee that corrections to prescribed roughness correlations vanish at $T=0$ in the limit $a\rightarrow 0$. }}
\label{countertermfig}
\end{figure}

 The 2-point correlation function of the roughness profile of\equ{d} decays exponentially at large momenta and the vertex functional given by\equ{Gammah} is quadratic in the field $\phi$. One-particle-irreducible (1PI) vertex functions with only external $\phi$-fields thus are finite if all 1PI vertices with only external roughness fields are. One therefore only requires counter terms for $n$-point vertex functions of the roughness profile. The 1-point counter term, $c^{(\eps)}_1$ is finite for $\eps\rightarrow 0^+$ and vanishes for $a\rightarrow\infty$. This counterterm is necessary for an unambiguous definition of the separation $a$. It ensures that\equ{novev} holds at all temperatures, separations and couplings. The parameter $a$ otherwise  would not always represent the mean separation of the plates. $c_1^{(\eps)}$ is the only counter term that depends on the plate separation $a$, temperature $T$ and \emph{both} coupling constants $\lambda$ and $\bar\lambda$. To leading order in the loop expansion, the equation $\vev{h}=0$ is shown diagrammatically in fig.~\ref{countertermfig} and gives,
\begin{align}
\label{c1}
\left. c^{(\eps=0)}_1(a;\lambda,\bar\lambda)\right|_\text{1-loop} &=-T\lambda\sum_n\int \frac{d\bk}{(2\pi)^2} g_{10}(\kappa_n)=- \frac{\partial}{\partial a}\frac{T}{4\pi}\sum_{n=-\infty}^\infty\int_{2\pi |n| T}^\infty \kappa d\kappa\ln(\Delta)\nonumber\\
&=-\frac{\partial}{\partial a}f^{(2)}(T;\lambda,\bar\lambda,a)\\
&\xrightarrow[{\lambda,\bar\lambda\sim\infty}]{}\  \ -\frac{\partial}{\partial a}\sum_{m=-\infty}^\infty\sum_{n=1}^\infty \frac{-a/\pi^2}{[(2 n a)^2+(m/T)^2]^2}\ \ \xrightarrow[{2 T a\ll 1}]{}\ \ -\frac{\pi^2}{480 a^4}\ ,\nonumber
\end{align}
where $f^{(2)}(T;\lambda,\bar\lambda,a)$ is the Casimir pressure at finite temperature on two semitransparent plates due to a massless scalar field of\equ{parallel0}. The last line in\equ{c1} reproduces the Casimir force on Dirichlet plates at finite\cite{Brown19691} and at zero temperature\cite{Casimir19481}. It is no coincidence that $c_1^{(\eps)}$ is the Casimir pressure since this counter term compensates for changes in the Casimir free energy due to  $\vev{h}\neq 0$. The correspondence is evidence for the correctness of the surface theory. Since it maintains $\vev{h}=0$,  the counterterm $c_1^{(\eps=0)}$ cancels all one-particle reducible contributions to the free energy.

Counter terms with more than one external roughness field ensure that prescribed correlation functions of the profile remain unchanged at $T=0$ when the two plates are (infinitely) far apart. These counter terms by definition depend only on the coupling $\lambda$ and on the cutoff $\eps$ and can  be computed using the fast parts of propagators in\equ{props} that survive the $a\rightarrow\infty$ limit. Since $g_{01}^{(f)}=0$, counter terms with an \emph{odd} number of external $h$-fields vanish in the limit $a\rightarrow\infty$. Apart from $c_1$ the model requires only counter terms $c^{(\eps)}_{2n}$ with an even number of external roughness profiles. To leading order in the loop expansion, $c^{(\eps)}_2(q;\lambda)$ is obtained by evaluating the diagrams of fig~\ref{countertermfig} at $T=0$ in the limit $a\rightarrow 0$. For $1/\lambda\gg \eps\rightarrow 0^+$ one obtains,
\begin{align}
\label{c2}
c^{(\eps)}_2(q;\lambda)&=\frac{\lambda^2}{2}\int_{-\infty}^\infty\frac{d\xi}{2\pi}\int \frac{d\bk}{(2\pi)^2} \frac{(\kappa-\kappa^\prime e^{-\eps\kappa^\prime}) e^{-\eps\kappa}}{2\kappa+\lambda}\ \ \text{with  }\ \begin{array}{lcl}
\kappa^2&=&\xi^2+\bk^2\\
\kappa^{\prime\,2}&=&\xi^2+(\bq-\bk)^2
\end{array} \nonumber\\
&=\frac{\lambda^2}{32\pi^2}\left[\frac{7}{\eps^3}-\frac{3\lambda}{2\eps^2}+\frac{3\lambda^2-q^2}{6\eps}+
\frac{q^2\lambda(23-24\gamma_E-24\ln(\eps\lambda))}{36} +\right.\\
&\left.\qquad\qquad+\frac{\lambda^3(1-3\ln 2)}{6}+\frac{5 q\lambda^2}{6}+\frac{q^3}{3}-
\frac{\lambda(\lambda+2 q)^3}{12 q}\ln(1+\frac{2q}{\lambda})\right]+{\cal O}(\lambda\eps)\ .\nonumber
\end{align}
For Dirichlet boundary conditions one must consider the strong coupling limit $0<1/\lambda\ll\eps\rightarrow 0^+$. The two-point counter term in this case simplifies to,
\bel{c2Dirichlet}
c^{(\eps)}_2(q;\infty)=\frac{\lambda}{2}\int_{-\infty}^\infty\frac{d\xi}{2\pi}\int \frac{d\bk}{(2\pi)^2} (\kappa-\kappa^\prime e^{-\eps\kappa^\prime}) e^{-\eps\kappa}=\frac{\lambda}{32\pi^2}\left[\frac{45}{\eps^4}+\frac{q^2}{6\eps^2}+\frac{q^4}{24}\right]+{\cal O}(q\eps)\ .
\ee
Together with the counter terms, the Feynman rules derived above define the loop expansion of this model. The total transverse momentum and thermal mode number are conserved at each vertex (assigning the time-independent $h$-field the Matsubara frequency $\xi_n=0$). This is a $2+1$-dimensional thermal field theory: the presence of another plate in a third spatial dimension manifests itself in the non-local dependence of propagators on the length scale "$a$".  From the point of view of the two-dimensional brane, this length scale could as well represent the Compton wave length of a massive particle. The model on the surface is holographic in the sense of\cite{Dvali20001,*Maldacena20051}.

\section{The Dirichlet (strong coupling) limit}
\label{Sec.Dirichlet}
The vertices in Eqs.~(\ref{oddv})~and~(\ref{evenv}) are all proportional to $\lambda$. To leading order in the strong coupling expansion, the propagators $g_{00},g_{01},g_{10}$ are of order $\lambda^{-1}$ and $g_{11}$ is of order $\lambda^0$. The leading superficial order, $N_\lambda$, of a Feynman diagram in the strong coupling regime thus is given by,
\bel{leadingorder}
N_\lambda=\#\gamma_{00}+\#\gamma_{01}+\#\gamma_{10}+\#\gamma_{11}-\#g_{00}-\#g_{01}-\#g_{10}\ ,
\ee
where $\#X$ denotes the number of $X$\rq{}s the diagram is composed of.
The quadratic model conserves the number of scalar surface fields and the $\psi$ and $\tpsi$ fields of propagators correspond to those of vertices.  \emph{Vacuum} diagrams thus satisfy the additional constraints,
\bal{constnr}
2\#\gamma_{00}+\#\gamma_{01}+\#\gamma_{10}&=&2\#g_{00}+\#g_{01}+\#g_{10}\nonumber\\
2\#\gamma_{11}+\#\gamma_{01}+\#\gamma_{10}&=&2\#g_{11}+\#g_{01}+\#g_{10}\nonumber\\
\#\gamma_{01}=\#\gamma_{10}&,&\#g_{01}=\#g_{10}\ .
\ea
Using\equ{constnr} in\equ{leadingorder}, the leading superficial order of a vacuum diagram is found to be given by the number of $g_{11}$ propagators it contains,
\bel{degreevac}
N_\lambda(\text{vac})=\#g_{11}=\#\gamma_{11}+\#g_{00}-\#\gamma_{00}\ .
\ee

If the strong coupling (Dirichlet) limit of the free energy is to exist, superficially divergent contributions with $N_\lambda(\text{vac})>0$ have to cancel. Such delicate cancellations generally arise due to underlying symmetries and are the consequence of associated Ward identities. A finite strong coupling limit for \emph{any} profile in this sense is a non-trivial condition on the surface model defined by \equ{Gammah}. That a Ward-like identity may ensure the existence of the strong coupling limit is suggested by the vertex functional $\Gamma^{(0)}$ for a flat plate given in\equ{eff0}. It evidently satisfies the identity,
\bel{ward0}
\frac{\delta}{\delta \tpsi_n(\bk)}\frac{\partial}{\partial \lambda}\Gamma^{(0)} = 0\ .
\ee
\equ{ward0} can be interpreted as stating that for vanishing profile the normal derivative $\tpsi$ need not vanish when Dirichlet boundary conditions are enforced. The original interaction with the profile by the $\delta$-function potential with the surface of\equ{interaction0} constrains the $\phi$-field at strong coupling but not its normal derivative. The strong coupling limit otherwise would not correspond to Dirichlet boundary conditions.  Even for non-vanishing profile, when $\psi$ and $\tpsi$ are coupled by $V^{01}[h]$, the strong coupling limit must not require \emph{both} surface fields to vanish.  One therefore expects an $h$-dependent linear combination of $\psi$ and $\tpsi$ to survive strong coupling and a generalization of\equ{eff0} to hold  for the vertex functional $\Gamma[\psi,\tpsi;h]$. Writing the linear combination of thermal modes in terms of an $h$-dependent functional $A_n[h]$, the generalization of \equ{ward0}  takes the form,
\bel{ward}
\left[\frac{\delta}{\delta \tpsi_n(\bk)}+\int\d{\bk^\prime}A_n(\bk,\bk^\prime;h)\frac{\delta}{\delta \psi_n(\bk^\prime)}\right]\frac{\partial}{\partial \lambda}\Gamma[\psi,\tpsi;h]= 0\ .
\ee
Inserting\equ{Gammah} in\equ{ward} and varying $\psi(\bk)$ and $\tpsi(\bk)$ leads to the two functional relations,
\bel{conditions}
A_n[h]\cdot(\one-V_n^{00}[h]) =V_n^{10}[h]\ \ \text{and}\ \ A_n[h]\cdot V_n^{01}[h]+V_n^{11}[h]=0\ .
\ee

A solution $A_n[h]$ to\equ{conditions} exists only if,
\bel{solution}
V^{11}_n[h]+V_n^{10}[h](\one-V_n^{00}[h])^{-1}V_n^{01}[h]=0 \ ,
\ee
for any profile $h$.  We have explicitly verified \equ{solution} to sixth order in the profile $h(\bk)$. Although we here do not provide a (combinatoric) proof of\equ{solution} to all orders, note that\equ{ward} would determine $A[h]$ and $V^{11}[h]$ for any choice of $V^{10}$ and $V^{00}$. Requiring that solutions to the wave equation with Dirichlet boundary conditions are not trivial determines the interaction $V_n^{11}[h]$ in terms of $V^{10}[h]$ and $V^{00}[h]$. \equ{solution} could also be considered to generate $\gamma_{11}$-vertices that are consistent with proposed $\gamma_{01}$ and $\gamma_{00}$ vertices. It is an indication of the consistency of the model that vertices with up to six external roughness fields satisfy\equ{solution}. We will not require higher vertices in our calculations and may safely assume that \equ{solution} in fact holds to all orders.
\begin{figure}
\includegraphics[width=\textwidth]{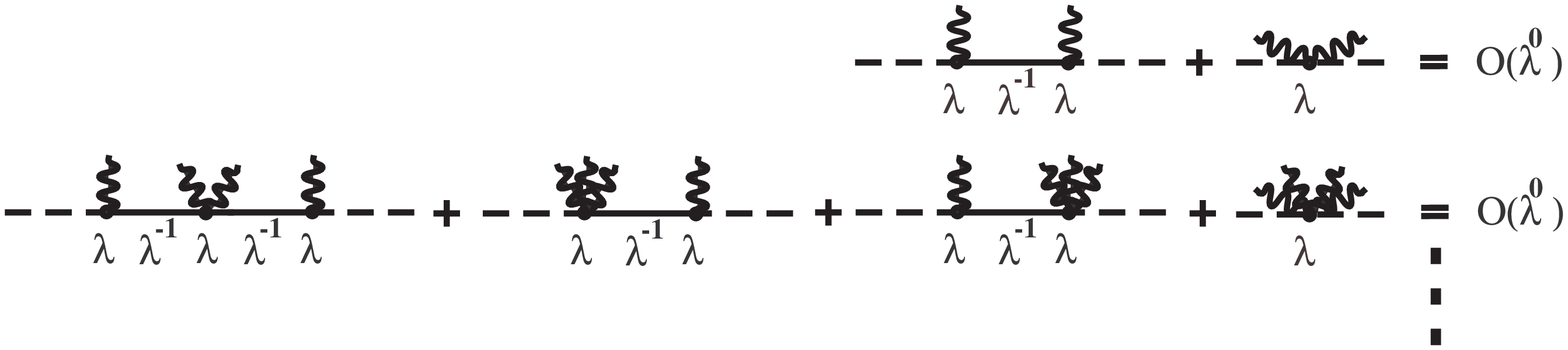}
\caption{Cancellation of the leading order in $\lambda$ of contributions to $\tpsi$-irreducible $\tpsi\tpsi$-vertices.  The solid lines represent $g_{00}$-propagators which at strong coupling are $\lambda^{-1}+{\cal O}(1)$.  $\tpsi$-irreducible $\tpsi\tpsi$-vertices thus are of leading superficial ${\cal O}(\lambda)$.  For the vertices of Eqs.~(\ref{evenv})~and~(\ref{oddv}), the leading superficial order cancels for any set of momenta and $\tpsi$-irreducible $\tpsi\tpsi$-vertices in fact are of ${\cal O}(1)$.}
\label{cancelfig}
\end{figure}

We still need to show that\equ{solution} is sufficient for a finite strong coupling limit of the effective action. In the following a connected Feynman diagram is $\tpsi$-reducible if it becomes disjoint by removing a single $g_{11}$ propagator and any number of $d$-propagators\footnote{A diagram that can only be separated by cutting $g_{00}$, $g_{01}$, $g_{10}$ propagators and any number of $h$-lines is $\tpsi$-irreducible. A one-particle reducible diagram thus can be $\tpsi$-irreducible.}.  In this quadratic model, a vertex is $\tpsi$-irreducible only if it contains no internal $g_{11}$ propagators. The analog of \equ{constnr} for a $\tpsi$-irreducible $\tpsi\tpsi$ vertex diagram with two external $\tpsi$-lines and no internal $g_{11}$ propagators implies that,
\begin{align}
2\#\gamma_{00}+\#\gamma_{01}+\#\gamma_{10}&=2\#g_{00}+\#g_{01}+\#g_{10}\nonumber\\
2\#\gamma_{11}+\#\gamma_{01}+\#\gamma_{10}&=2+\#g_{01}+\#g_{10}\nonumber\ .
\end{align}
Its leading superficial order in $\lambda$ therefore is,
\begin{align}
\label{ordervertex}
N_\lambda(\tpsi\text{-irred. } \tpsi\tpsi\text{-vertex})&=\#\gamma_{00}+\#\gamma_{01}+\#\gamma_{10}+\#\gamma_{11}-\#g_{00}-\#g_{01}-\#g_{10}\nonumber\\
&=\#\gamma_{11}+\#g_{00}-\#\gamma_{00}=1\ .
\end{align}
Eq.~(\ref{solution}) on the other hand implies that the leading order in $\lambda$ of all contributions to an $\tpsi$-irreducible $\tpsi\tpsi$-vertex in fact cancels. The superficial order in $\lambda$ in\equ{ordervertex} does not account for this cancellation among contributions of the same superficial order and a $\tpsi$-irreducible $\tpsi\tpsi$-vertex therefore is at most of order $\lambda^0$.

The superficial order in the coupling $\lambda$ of a vacuum diagram was found to be just $\#g_{11}$ in\equ{degreevac}, because this is precisely the number of $\tpsi$-irreducible $\tpsi\tpsi$-vertices the diagram contains. Since we have just seen that \equ{solution} implies that a $\tpsi$-irreducible $\tpsi\tpsi$-vertex in fact contributes at most is ${\cal O}(1)$, the combined contribution to the free energy of all vacuum diagrams with a given number of $g_{11}$ propagators also is at most of order ${\cal O}(1)$. \equ{ward} thus  ensures a finite free energy in the strong coupling (Dirichlet) limit.

\section{Two-loop contribution to the free energy: the leading roughness correction}
\label{leadingcorr}
\begin{figure}
\includegraphics[scale=0.50]{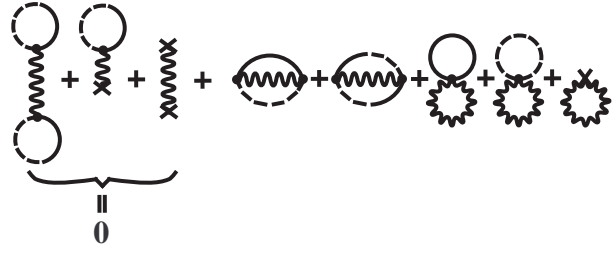}
\caption{Two-loop vacuum diagrams}
\label{twoloopvac}
\end{figure}
The two-loop vacuum diagrams of Fig.~\ref{twoloopvac} give the leading roughness correction to the free-energy. The evaluation simplifies and is more transparent in the Dirichlet limit for both plates. The correction to the Casimir free energy per unit area of a massless scalar field for two parallel plates due to the roughness of one of them in this strong coupling limit is given by the rather simple expression,
\begin{align}
\label{twoloopDir}
\Delta f_\text{D}^{(2)}(\sigma,\ell;a,T)=-T\sum_n\int\frac{d\bk d\bk^\prime}{(2\pi)^4} \frac{\kappa_n\kappa_n^\prime\,d(\bk-\bk^\prime)}{(e^{2a\kappa_n}-1)(1-e^{-2a\kappa_n^\prime})}\ ,
\end{align}
where $d(\bk)$ is the two-point correlation function of the roughness profile.  There in addition is a finite correction to the free energy due to the roughness for an individual plate. It does not depend on the separation $a$ and therefore does not lead to a modification of the force on the plate and will be ignored here. The correction to the interaction of\equ{twoloopDir} depends on the exact form of $d(\bk)$, but some conclusions about its general behavior can be drawn in the limit of large, $\ell\gg a$, and of small, $\ell\ll a$, correlation length. For $\ell\gg a$, the support of the roughness correlation $d(\bk-\bk^\prime)$, such as the one of \equ{d}, is restricted to $|\bk-\bk^\prime| a\ll 1$. One thus may replace $\kappa^\prime$ by $\kappa$ in the integrand without great error. With  $\sigma^2=(2\pi)^{-2}\int d\bk d(\bk)$ this gives the universal limit,
\begin{align}
\label{pfalimit}
\Delta f_\text{D}^{(2)}(\sigma,\ell\gg a,T)&\sim-\frac{\sigma^2}{2}\left(\frac{\partial}{\partial a}\right)^2f^{(2)}(T;\infty,\infty,a)\\
&\sim\frac{\sigma^2}{2}\frac{\partial^2}{\partial a^2}\sum_{m=-\infty}^\infty\sum_{n=1}^\infty \frac{-a/\pi^2}{[(2 n a)^2+(m/T)^2]^2}\ \ \xrightarrow[{2 T a\ll 1}]{}\ \ -\frac{\pi^2\sigma^2}{240 a^5}\ ,\nonumber
\end{align}
which does not depend on the specific form of the correlation function $d(\bk)$. As should be expected\cite{Blocki19771},\equ{pfalimit} coincides with the  roughness correction in PFA for profiles with large correlation length\cite{Klimchitskaya19991,Bordag2009bk}.
\begin{figure}
\begin{minipage}{\textwidth}
\begin{minipage}{0.49\textwidth}
\includegraphics[width=\textwidth]{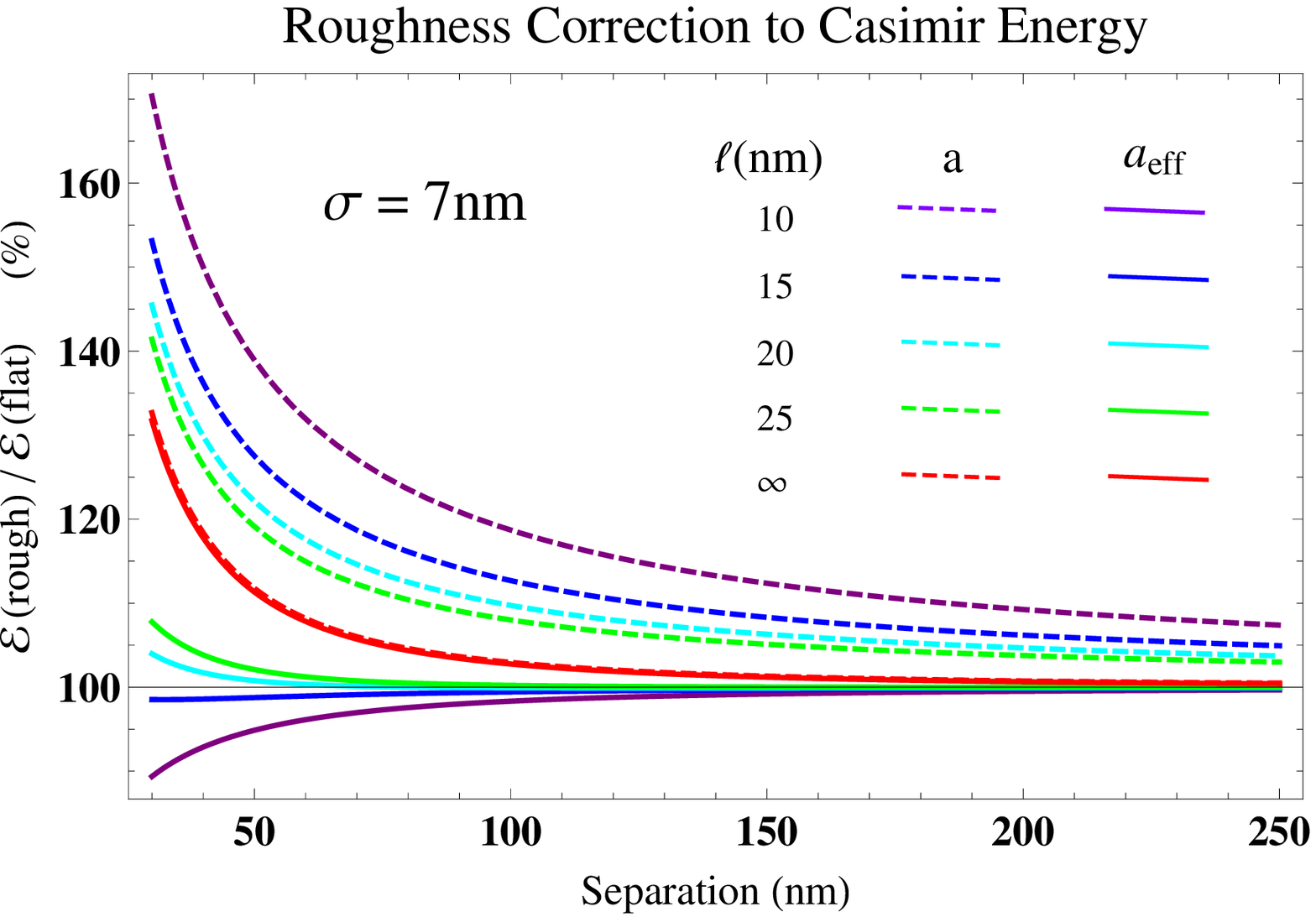}
\label{PertEnergy}
\end{minipage}
\begin{minipage}{0.49\textwidth}
\includegraphics[width=\textwidth]{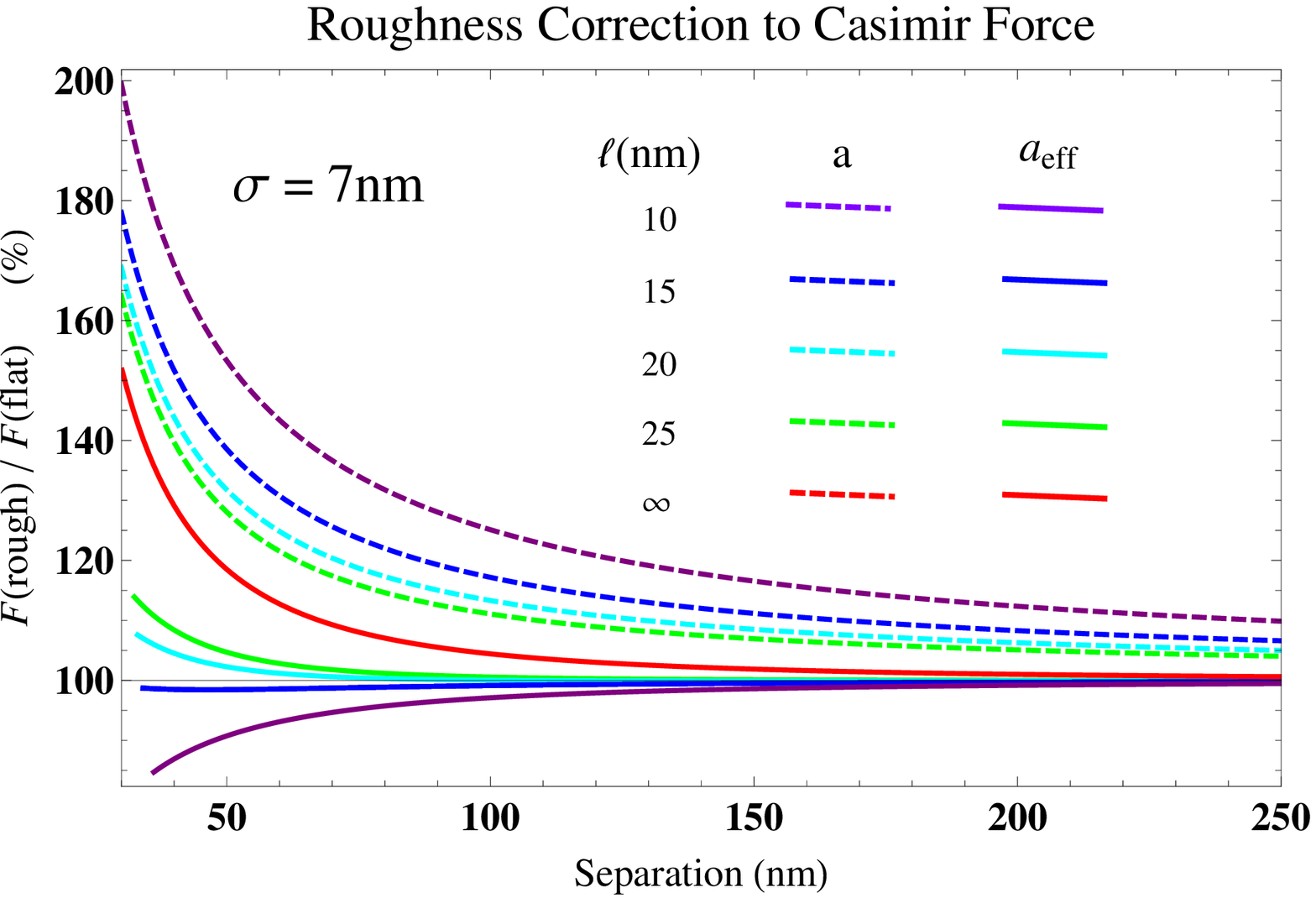}
\label{PertForce}
\end{minipage}
\end{minipage}
\caption{{\small (color online) Relative roughness corrections to the Casimir energy and Casimir force in \% due to a scalar satisfying Dirichlet boundary conditions on two plates, one of which is flat, the profile of the other is characterized by its variance $\sigma^2=49$nm$^2$ and correlation length $\ell$. In two-loop approximation the correction is proportional to $\sigma^2$. Pairs of dashed and solid curves of the same color correspond to the same $\ell=10\text{nm (violet)},15\text{nm (blue)}, 20\text{nm (cyan)}, 25\text{nm (green)  and } \ell=\infty$  (from the outer pair of curves to the inner). Dashed curves represent the correction as a function of the mean separation $a$, whereas solid curves show it as a function of the effective separation $a_{\text{eff}}=a-\frac{\sigma^2}{\ell}\sqrt{\frac{\pi}{2}}$. The (red) PFA correction for $\ell=\infty$ is the same in both cases. }}
\label{PertCorrections}
\end{figure}
In the opposite limit of short correlation length, $\ell\ll a$, or at large separations,  the  $\bk$-integral is exponentially restricted to the domain $|\bk|\lesssim 1/a$ whereas the $\bk^\prime$-integral in\equ{twoloopDir} is finite only because roughness correlations are negligible for $|\bk^\prime|\gg 1/\ell\gg 1/a$. The leading behavior of the roughness correction at separations $a\gg\ell$ thus is,
\begin{align}
\label{twoloopDir0}
\Delta f^{(2)}(\ell\ll a,\lambda\sim\infty)&\sim-\int\frac{d\bk^\prime}{(2\pi)^2} k^\prime\,d(\bk^\prime)\times T\sum_n\int\frac{d\bk}{(2\pi)^2} \frac{\kappa_n}{(e^{2a\kappa_n}-1)}\nonumber\\
&=-\left(\frac{\sigma^2}{\ell}\sqrt\frac{\pi}{2}\right)\times\frac{\partial}{\partial a} f^{(2)}(T;\infty,\infty,a)\\
&=f^{(2)}(T;\infty,\infty,a^D_{\rm eff})-f^{(2)}(T;\infty,\infty,a)+{\cal O}\left(\frac{\sigma^4}{a^5\ell^2}\right)\ ,\nonumber
\end{align}
where $f^{(2)}(T;\infty,\infty,a)$ is the free energy of\equ{parallel0} for two flat parallel Dirichlet planes at separation $a$ and
\bel{aeff}
a^D_\text{eff}=a-\int\frac{d\bk}{(2\pi)^2} k\,d(\bk)\sim a-\frac{\sigma^2}{\ell}\sqrt{\frac{\pi}{2}}\ .
\ee
The shift away from the mean of the profile is always of order $\sigma^2/\ell$, but the proportionality constant depends somewhat on the shape of the correlation function $d(\bk)$ and is $\sqrt{\pi/2}\sim 1.25\dots$ for the one of\equ{d} only.  Note that this displacement in the apparent surface of the profile is within the "thickness" of the profile for  $\sigma<\ell$. This mild condition generally is satisfied by naturally rough surfaces that arise from random dislocations of surface atoms and is a requirement for the validity of the loop expansion. However,  it should be noted that surfaces with $\sigma>\ell$ can be artificially created. In this case a loop expansion of the free energy in $\sigma^2/\ell^2$ is not applicable\cite{Chan20081,Lambrecht20081,*Lambrecht20101}.

The perturbative roughness correction is well known\cite{Maradudin19801,*Mazur19811,Novikov19901,*Novikov19902,*Novikov19921,*Novikov19922,
Genet20031,Neto20051,*Neto20061, Zwol20071,*Zwol20081,*Zwol20082, Bordag2009bk}. A systematic loop expansion that includes temperature and roughness corrections simultaneously and yields a relatively simple closed expression as in\equ{twoloopDir} to our knowledge was not considered previously. Although it may not be worth the effort for the scalar field theory, the systematic inclusion of higher orders in the loop expansion in principle is straightforward in this field theoretic approach.

 Although the shift in\equ{aeff} generally is quite small and well within the profile\rq{}s thickness, the effect on the roughness correction can be dramatic. As shown in\fig{PertCorrections}, or as can be deduced by examining\equ{twoloopDir}, the perturbative roughness correction tends to \emph{increase} with \emph{decreasing} correlation length when the mean separation between the two plates is used as reference. The correction is quite large even for $a\gg \ell $ and easily exceeds $20\%$ at experimentally accessible separations for typical roughness profiles\cite{Zwol20071,*Zwol20081,*Zwol20082}. However, this effect to a great extent is eliminated by redefining the effective planar \lq{}surface\rq{} of a rough plate. As shown in \fig{PertCorrections}, the residual roughness correction \emph{decreases} with \emph{decreasing} correlation length $\ell$ if the effective separation $a^D_{\text{eff}}$ of\equ{aeff} is used for the separation. Thus, at least to leading order in the loop expansion, the main effect of roughness is to define the reference plane of the plate. This reference plane generally is \emph{not} the mean of the profile.

To determine the absolute separation of rough plates can be experimentally challenging. The previous considerations suggest that one could instead experimentally calibrate the (effective) separation of two plates so as to eliminate asymptotic $1/a^4$ corrections to the Casimir interaction energy of flat parallel plates (or asymptotic $1/a^5$-corrections to the force). In terms of this more appropriate definition of the separation, the leading asymptotic correction to the force for large $a\gg \ell$ is of order $1/a^6$ only. Note that PFA-corrections, corresponding to infinite correlation length, are of this order and are not changed by this procedure. Such an intrinsic determination of the effective absolute separation $a^D_{\text{eff}}$  eliminates systematic errors due to electrostatic and other means of deducing the average separation of rough surfaces and facilitates a theoretical interpretation of the experiments. However, the suggested asymptotic calibration suffers from the fact that the Casimir force and therefore the signal-to-noise ratio decrease rapidly with increasing separation. A truly asymptotic determination is impractical and a compromise is required. \fig{PertCorrections} suggests that intermediate separations $(100\text{nm}<a<300\text{nm})$ could be used to optimize this procedure in most experimental situations. In terms of the thus optimized separation, corrections to the Casimir force of two flat plates are much smaller and under better theoretical control.

As argued in the introduction, an improved definition of the effective separation is necessary to avoid the conclusion that the reflection coefficient of a rough plate at long wavelengths ($a\gg\ell$) is larger than for a perfectly reflecting mirror and that unitary is violated. With a correct definition, the scattering matrix of a rough plate and the corresponding Casimir force ought to both \emph{decrease} in magnitude compared to those of a flat (perfectly reflecting) Dirichlet plate. One furthermore expects the scattering matrix and Casimir force to \emph{decrease} in magnitude with decreasing $\ell$ for $a\gg\ell$. Both physical requirements are met for $a\gg\ell$ by using the effective separation $a^D_{\text{eff}}$ defined in\equ{aeff}. The corresponding force is always weaker than the PFA suggests. We now show that an appropiate definition of the absolute separation to a rough plate leads to a scattering matrix with physically acceptable properties in the limit $\ell\gg a$.

\section{The limit $a\gg\ell$: An effective low-energy field theory.}
In the limit $\ell\ll a$ the two-point correlation function $D_2(\bx)$ of the profile is \emph{localized} to $|\bx|\lesssim \ell\ll a$ and we can use renormalization group techniques\cite{Weinberg2005bk} to analyze the situation. In this limit we can approximately "integrate out" high momentum contributions and construct an effective theory of surface fields for wave numbers $|\bk|\lesssim 1/a$. In our model, the separation of momentum scales already occurs for tree-level surface field propagators.  In\equ{props} they naturally decompose into $(f)$ast and $(s)$oft components.

A local vertex $v_{N \tilde N}$ of the effective low-energy theory corresponds to the sum of all connected diagrams that contain only $(f)$ast internal propagators and have $2N$ $(s)$oft external $\psi$- and $2\tilde N$ $(s)$oft external $\tilde\psi$-fields and no external $h$-fields. Since the range of the roughness correlation $D_2(\bx-\by)$ vanishes, these vertices are local  in the limit $\ell\rightarrow 0$ and by construction do not depend on the presence of another plate at separation $a$ and coupling $\bar\lambda$. Because $g^{(f)}_{01}=0$, the effective local vertices furthermore do not mix dynamical surface fields and conserve the number of $\psi$- and $\tpsi$- fields individually. They vanish unless the number of external $\psi$ and $\tpsi$ fields are \emph{both} even. Closed loops of fast surface fields correspond to separation-independent corrections to roughness-correlations that are precisely canceled by the corresponding counter-terms at $T=0$. At low temperatures it therefore suffices to consider connected vertex diagrams with only fast internal propagators and no closed internal loops of dynamical surface field propagators. The effective local vertices are finite since all loop momenta are restricted to $|\bk|\ell\lesssim 1$ by the roughness-correlations, but some tend to diverge for $\ell\rightarrow 0$. To determine the degree of divergence with $\ell$, note that the number, $L$, of transverse momentum loops of a connected vertex diagram with $2(N+\tilde N)$ external surface fields is given by,
\bel{nloops}
L=\#d+1-N-\tilde N
\ee
where $\#d$ is the number of roughness-propagators $d(\bk)$ the diagram contains. The canonical mass-dimension $[v_{N\tilde N}]$ of a ${N\tilde N}$-vertex in transverse momentum space is,
\bel{dimv}
[\lambda^{-2 N} v_{N\tilde N}]=2-3N-3\tilde N\ .
\ee
In\equ{dimv} a factor of $\lambda^{-1}$ provided by  $g^{(s)}_{00}$ and $g^{(s)}_{01}$ propagators in vacuum diagrams was included for each external $\psi$-field. We distinguish two extreme limits:

\begin{figure}
\includegraphics[scale=0.50]{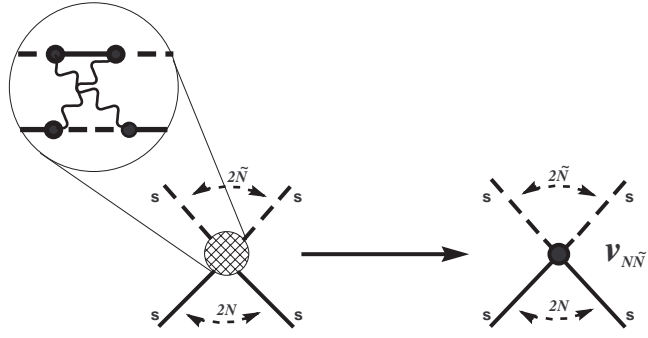}
\caption{Localization of connected vertices in the $\ell\rightarrow 0$ limit. Only $(f)$ast components of internal dynamical surface propagators contribute. External momenta are limited to $|\bk|<1/a\ll 1/\ell$.  }
\label{Localization}
\end{figure}

\subsection{A rough Dirichlet plate: $1/\lambda\ll \ell\ll a$}
For $1/\lambda\ll \ell$ the internal $g^{(f)}_{00}$ propagators (given in\equ{g00}) of an effective local vertex may be  approximated by $1/\lambda$. $\lambda\ell\gg 1$ includes the case of Dirichlet boundary conditions on the rough plate and we  for simplicity consider only this extreme limit. The leading contribution to an $N\tilde N$-vertex is proportional to $\lambda^N$ due to the cancellations observed in Sec.~\ref{Sec.Dirichlet}. Note that internal $d$-propagators of a diagram are proportional to $\sigma^2$.  For vanishing external momenta Eqs.~(\ref{nloops})~and~(\ref{dimv}) then imply that the sum of $L$-loop contributions to the effective local vertex behaves as,
\bel{behaveDir}
\lambda^{-2 N} v^{(L)}_{N\tilde N}\propto \frac{\sigma^2}{\ell}\left(\sigma^2\ell\right)^{N+\tilde N-1}\left(\frac{\sigma^2}{\ell^2}\right)^{L-1}\left(1+{\cal O}\left((\lambda\ell)^{-1}\right)\right)\ ,
\ee
in the strong coupling limit. Upon summing the $\#L$-loop contributions, the local effective vertex at vanishing external momentum in the Dirichlet limit thus is of the form,
\bel{vNN}
v^D_{N\tilde N}=\sum_{L=1}^\infty v^{(L)}_{N\tilde N}=\lambda^{2 N}\frac{\sigma^2}{\ell}\left(\sigma^2\ell\right)^{N+\tilde N-1}Q^D_{N\tilde N} \left(\frac{\sigma^2}{\ell^2}\right)\ ,
\ee
where the dimensionless functions $Q^D_{N\tilde N}(s)$ are analytic at $s=0$.  We emphasize that the effective vertices reflect properties of the rough plate only. They do not depend on properties of the other parallel plate and we have the desired separation of scales. $\sigma^2/\ell^2<0.1$ for typical surfaces used in Casimir studies\cite{Decca20031,*Decca20071,Zwol20071,*Zwol20081,*Zwol20082} and low orders in the loop expansion should provide fairly accurate local vertices $v^D_{N\tilde N}$ in the strong coupling limit.

Before proceeding to evaluate local effective vertices to leading order in the (hard) loop expansion, observe that the function $Q^D_{N\tilde N}$ in\equ{vNN} depends on the two-point correlation function $d(\bk)$ and, in principle, also depend on higher correlation functions  of the roughness profile. It therefore largely is a matter of perspective whether the vertices $v_{N\tilde N}$ of\equ{vNN} or the correlation functions $D_n$ of\equ{def-D} are considered as the primary parameters that describe the plates roughness in the low energy effective field theory. Of course, not every set of local vertices $v^D_{N\tilde N}$ corresponds to a physically realizable profile. A model of the correlations provides a basis for appropriate values and relations among the effective vertices $v^D_{N\tilde N}$ of the low-energy description. Nevertheless, within a certain domain, the phenomenological parameters of the low-energy effective theory in effect are its local vertices. Assuming an analytic continuation of the functions $Q^D_{N\tilde N} (s)$ to $s>1$ to exist, this effective low-energy description in the limit $a\gg\ell$ can be extended to a region of the parameter space where the loop-expansion is no longer valid\cite{Chan20081,Lambrecht20081,*Lambrecht20101}.   

Observe that the dependence of effective local vertices on external momenta of the $(s)$oft fields gives rise to  contributions to the free energy that are suppressed by powers of $\ell/a$. To leading order in the loop expansion in $\sigma^2/\ell^2$, \equ{vNN} implies that effective local vertices with more than four $(s)$oft external fields can be ignored in the limit $\ell/a\to 0$. As in chiral perturbation theory\cite{Gasser19841} one arrives at an expansion in the canonical dimension of local vertices,  vertices with more external fields becoming  relevant at higher orders of the (soft) loop expansion only.  Allowing for at most one hard internal loop,  the present low-energy effective model (with the correlation function $d(\bk)$ of\equ{d}) has the following local effective vertices,
\begin{subequations}
\label{v1loop}
\begin{align}
v^D_{01}&=\ell^2\sigma^2\int_0^\infty \hspace{-1em}k dk\left[\frac{\lambda^2}{2 k+\lambda}-\lambda\right]e^{-k^2 \ell^2/2}&\xrightarrow[\lambda\ell\gg 1]{}&-\frac{\sigma^2}{\ell}\sqrt{2\pi}\delta_{\bar n\bar n^\prime}\label{v01}\\
v^D_{10}&=-\lambda^2 \ell^2\sigma^2\int_0^\infty\hspace{-1em} k dk \frac{k}{2} e^{-k^2 \ell^2/2}&\xrightarrow[\lambda\ell\gg 1]{}&-\frac{\lambda^2\sigma^2}{4\ell}\sqrt{2\pi}\delta_{nn^\prime}\label{v10}\\
v^D_{02}&=2\pi\ell^4\sigma^4\int_0^\infty\hspace{-1em} k dk\left[\frac{2\lambda^4}{(2 k+\lambda)^2}-\frac{4\lambda^3}{2 k+\lambda}+2\lambda^2\right]e^{-k^2 \ell^2}&\xrightarrow[\lambda\ell\gg 1]{}&8\pi\sigma^4\delta_{\bar n\bar n^\prime}\delta_{\bar m\bar m^\prime}\label{v02}\\
v^D_{11}&=2\pi\lambda^2\ell^2 \sigma^2+2\pi\ell^4\sigma^4\int_0^\infty\hspace{-1em} k dk\left[\lambda^3 k-\frac{\lambda^4 k}{2 k+\lambda}\right]e^{-k^2 \ell^2}&\xrightarrow[\lambda\ell\gg 1]{}& 2\pi\lambda^2\sigma^2(\ell^2\delta_{n\bar n}\delta_{m\bar m}+\sigma^2\delta_{nm}\delta_{\bar n \bar m})\label{v11}\\
v^D_{20}&=2\pi\ell^4\sigma^4\int_0^\infty \hspace{-1em}k dk 2\left(\frac{\lambda^2 k}{2}\right)^2 e^{-k^2 \ell^2}&\xrightarrow[\lambda\ell\gg 1]{}&\quad \frac{\pi\lambda^4\sigma^4}{2}\delta_{nn^\prime}\delta_{mm^\prime}\label{v20}\\
\text{etc.}&&&\nonumber
\end{align}
\end{subequations}
Only the final expressions of\equ{v1loop} include the Kronecker symbols giving the conservation of Matsubara modes [indices with a bar designate $\tpsi$ modes]. Note that the tree-level ("one-roughness-exchange") and one-loop contributions to $v_{11}$ differ in their flow of mode indices. The individual terms  in the intermediate expressions correspond to diagrams of different topology that contribute to the effective vertex.

The free energy of the effective low-energy theory with the $(s)$oft propagators of\equ{props} and local vertices of\equ{v1loop} describes separation-dependent corrections due to the profile of a Dirichlet plate at separations $a\gg \ell$ from a smooth parallel plate. The free energy of the low-energy effective theory in powers of $\sigma/a$ may be obtained in the $(s)$oft loop expansion. The effective 2-point vertices $v_{01}$ and $v_{10}$ play a crucial r{\^o}le in this expansion. They correct the low-energy behavior of propagators and thus affect all higher orders of the $(s)$low loop expansion as well.

\begin{figure}
\includegraphics[scale=0.50]{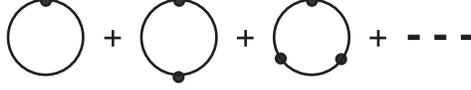}
\caption{One loop contributions to the free energy in the effective model for a rough Dirichlet plate.}
\label{oneloopEffective}
\end{figure}

The ratio  $v^D_{10}:v^D_{01}=\lambda^2/4$ precisely compensates for the ratios of soft propagators $g^{(s)}_{11}:g^{(s)}_{10}:g^{(s)}_{00}=-\lambda/(2t)$ in the Dirichlet ($t\rightarrow t_D=1$) limit. The roughness and separation-dependent correction, $\Delta f^{(2)}$, to the free energy per unit area of the effective low energy model thus is obtained by evaluating the 1-loop diagrams of \fig{oneloopEffective} with an effective interaction $-2\rho^D=2 v^D_{01}$ and propagator $g^{(s)}_{11}$,
\begin{align}
\label{effD}
\Delta f^{(2)}_\text{1-loop}(\rho^D,a\gg\ell\gg 1/\lambda) &= -T\sum_n\int\frac{d\bk}{(2\pi)^2}\sum_{k=1}^\infty\frac{1}{2 k} \left(2\rho^D \frac{\kappa_n\bar t_n e^{-2\kappa_n a}}{2\Delta_n}\right)^k\nonumber\\
&=\frac{T}{2}\sum_n\int\frac{d\bk}{(2\pi)^2}\left[\ln[\Delta_n-\rho^D\kappa_n\bar t_n e^{-2\kappa_n a})-\ln(\Delta_n)\right]\nonumber\\
&=\frac{T}{2}\sum_n\int\frac{d\bk}{(2\pi)^2}\left[\ln(1-(1+\rho^D\kappa_n)\bar t_n e^{-2\kappa_n a})-\ln(\Delta_n)\right]\nonumber\\
&=\frac{T}{2}\sum_n\int\frac{d\bk}{(2\pi)^2}\left[\ln(1-t^D_\text{rough}(\kappa_n)\bar t_n e^{-2\kappa_n a^D_\text{eff}})-\ln(\Delta_n)\right]
\end{align}
The one-loop free energy depends on the mean plate separation $a$ and on a length $\rho^D=-v_{01}\sim\sqrt{2\pi}\sigma^2/\ell$ that characterizes the profile. The effective separation of the two plates, $a^D_\text{eff}=a-\rho^D/2$, in one-hard-loop approximation coincides with the one found perturbatively in\equ{aeff}. We in addition 
obtain the reduced scattering matrix $t^D_\text{rough}$ for low-energy scattering off a rough Dirichlet plate,
\bel{teffD}
t^D_\text{rough}(\kappa)=(1+\rho^D\kappa)e^{-\rho^D\kappa}\ .
\ee
$t^D_\text{rough}(\kappa)$ is positive and never exceeds unity. It satisfies all the requirements of a reduced scattering matrix and is consistent with phenomenology for scattering off a rough plate in that only short wavelengths with $\kappa\rho^D\gg 1$ are strongly affected. $t^D_\text{rough}(\kappa)<1$ is due to diffuse scattering of part of the incident wave with (transverse) wave vector $\bk$. The intensity of the outgoing wave with (transverse) wave-vector $\bk$ is thereby reduced. Diffuse scattering by a rough surface is more effective at shorter wavelengths and negligible for wavelengths that are long compared to $\rho^D\sim \sigma^2/\ell$. It perhaps is worth remarking that the scattering matrix found in this approximation does not depend on the separation $a$ of the two plates (as the $GTGT$-formula\cite{Kenneth20061} indeed requires). However, our approximations in deriving the low energy effective theory can only be justified for $a\gg\ell$.

\subsection{A rough semi-transparent plate: $a\gg\ell\ll 1/\lambda$}
This limit includes that of weak coupling. We proceed similarly as for the Dirichlet case but $\ell$ now is the smallest correlation length.  The leading behavior of a local vertex thus is determined by its degree of divergence as the \lq{}cutoff\rq{} $\ell$ on hard loop momenta is removed. In the limit of large transverse momenta we have that $g_{00}^{(f)}\sim 1/k, g_{11}^{(f)}\sim-k/2$. Neither depends on $\lambda$ and a local vertex in this case is proportional to $\lambda^{N_V}$, where $N_V$ is the total number of primitive vertices it is composed of. Eqs.~(\ref{dimv})~and~(\ref{nloops}) then imply that for vanishing external momenta,
\bel{behaveSemi}
v^{(\#L)}_{N\tilde N}\propto \frac{\lambda^{2 N}}{\ell^2}(\ell\sigma^2)^{\tilde N+N}(\lambda\ell)^{N_V-2 N}\left(\frac{\sigma^2}{\ell^2}\right)^{\#L-1}\left(1+{\cal O}\left(\lambda\ell\right)\right)\ .
\ee
The meaning of\equ{behaveSemi} becomes more transparent upon noting that for any connected diagram $N_V-2 N\ge \tilde N-N+1$. At any given order in the loop expansion,  the largest contribution to an effective local vertex in the present limit arises from diagrams with the minimal number $N_V=\tilde N+N+1$ of internal vertices. Local $n$-point vertices with external $\tpsi$-fields thus are suppressed by powers of $\lambda\ell\ll 1$ compared to vertices that describe the scattering of slow $\psi$-quanta only. In the present limit, the low energy effective   model is described by a scalar surface field with propagator $g^{(s)}_{00}$ and local $v_{N0}$ vertices only. To first order in the loop expansion one again obtains the vertices of Eqs.~(\ref{v10})~and~(\ref{v20}),
\begin{subequations}
\label{v1loopS}
\begin{align}
v_{10}&=-\lambda^2 \ell^2\sigma^2\int_0^\infty\hspace{-1em} k dk \frac{k}{2} e^{-k^2 \ell^2/2}&\xrightarrow[\lambda\ell\ll 1]{}&-\frac{\lambda^2\sigma^2}{4\ell}\sqrt{2\pi}\delta_{nn^\prime}\label{v10s}\\
v_{20}&=2\pi\ell^4\sigma^4\int_0^\infty \hspace{-1em}k dk 2\left(\frac{\lambda^2 k}{2}\right)^2 e^{-k^2 \ell^2}&\xrightarrow[\lambda\ell\ll 1]{}&\quad \frac{\pi\lambda^4\sigma^4}{2}\delta_{nn^\prime}\delta_{mm^\prime}\label{v20s}\ ,\\
\end{align}
\end{subequations}
but the local effective interactions $v_{01},v_{11}$ and $v_{02}$ now are negligible. To one $(s)$oft loop, the roughness and separation-dependent correction $\Delta f^{(2)}$ to the free energy per unit area in the limit $a\gg\ell\ll 1/\lambda$ is,
\begin{align}
\label{effS}
\Delta f^{(2)}_\text{1-loop}(\rho;a\gg\ell\ll 1/\lambda ) &= -T\sum_n\int\frac{d\bk}{(2\pi)^2}\sum_{k=1}^\infty\frac{1}{2 k} \left(\rho \frac{\kappa_n t_n\bar t_n e^{-2\kappa_n a}}{\Delta_n}\right)^k\nonumber\\
&=\frac{T}{2}\sum_n\int\frac{d\bk}{(2\pi)^2}\left[\ln(\Delta_n- \rho\kappa_n t_n\bar t_n e^{-2\kappa_n a})-\ln(\Delta_n)\right]\nonumber\\
&=\frac{T}{2}\sum_n\int\frac{d\bk}{(2\pi)^2}\left[\ln(1-(1+\rho\kappa_n)t_n\bar t_n e^{-2\kappa_n a})-\ln(\Delta_n)\right]\nonumber\\
&=\frac{T}{2}\sum_n\int\frac{d\bk}{(2\pi)^2}\left[\ln(1-t_\text{rough}(\kappa_n)\bar t_n e^{-2\kappa_n a_\text{eff}})-\ln(\Delta_n)\right]\ ,
\end{align}
with
\bel{as}
t_\text{rough}(\kappa)=(1+\kappa\rho) t(\kappa) e^{-\kappa\rho}\le t(\kappa) \text{   and   }  a_\text{eff}=a-\rho/2 \ .
\ee
One  reproduces the Dirichlet-plate result of\equ{teffD} by simply setting $t=t^D=1$ and $\rho=\rho^D$ in\equ{as},  but for the same profile the parameter $\rho$ is only half that found in the Dirichlet limit,
\bel{alph}
\rho=\frac{\sigma^2}{\ell}\sqrt{\frac{\pi}{2}}=\rho^D/2\ .
\ee
The displacement of the equivalent surface of the rough plate from the mean of the profile by $\rho$ evidently also depends on the transparency of the plate.  Considering the effective shift $\rho(\lambda)$ as a phenomenological parameter of the rough plate, the main effect due to roughness in the limit $a\gg \ell$ is to define the position of the planar scattering surface and simultaneously modify the scattering matrix of a flat plate as in\equ{as}. It is interesting that the modification of the scattering matrix and the shift from the mean of the profile are not independent. The two extreme limits we considered provide a range for the parameter $\rho(\lambda)$ in terms of the variance and correlation length of a profile described by\equ{d},
\bel{limits}
\sqrt{\frac{\pi}{2}}\frac{\sigma^2}{\ell}\le\rho(\lambda)\le\sqrt{2\pi}\frac{\sigma^2}{\ell}\ .
\ee
The upper bound of\equ{limits} corresponds to a rough surface with Dirichlet boundary conditions and the lower to weak coupling. Note that the effective scattering plane does not coincide with the mean of the profile even for weak coupling $\lambda\sim 0$ [although the scattering matrix is arbitrary small].

\section{Discussion}
We have developed a field theoretical approach to the Casimir free energy of a massless scalar field in the presence of parallel rough and smooth semi-transparent plates. Changes in the free energy due to interaction of the scalar with the rough surface were found to be described by an effective $2+1$-dimensional field theory on the equivalent plane involving two dynamical surface fields, $\psi$ and $\tpsi$ as well as the static profile $h$.   The model on this planar boundary of the original space is holographic in that the existence of another dimension and of a second parallel plate at a separation $a$ are encoded in non-local propagators. The theory in this sense is a low-dimensional analog of brane models in string theory\cite{Dvali20001,*Maldacena20051}.

Two-loop contributions to the free energy of this model give the leading roughness correction. This leading  correction for a massless scalar field is qualitatively similar to that obtained by perturbative analysis for electromagnetic fields\cite{Maradudin19801,Novikov19901,Genet20031,*Neto20051,*Neto20061,Zwol20071}, but its field theoretic origin allows for a consistent inclusion of finite temperature effects and for a more transparent interpretation. For a scalar field in the strong coupling (Dirichlet) limit, the leading loop correction was obtained  in closed form in \equ{twoloopDir} and is shown in fig.~\ref{PertCorrections}. As for the electrodynamic corrections considered in\cite{Genet20031,Neto20051,*Neto20061,Zwol20071,*Zwol20081,*Zwol20082} the PFA result \cite{Klimchitskaya19991,Bordag2009bk} is reproduced for $a\ll\ell$ and the Casimir force appears to strengthen with decreasing $\ell/a$. We argued in the introduction that this apparently violates unitarity when $a\gg\ell$. 

The problem can be traced to an inappropriate choice of the equivalent planar surface for a rough plate. This plane does not coincide with the mean of the profile but is displaced a distance $\rho\propto \sigma^2/\ell$ from it. With the correct definition of the effective surface, roughness corrections are much smaller and  the Casimir force \emph{weakens} with increasing roughness $\sigma^2/\ell$. Roughness strengthens the Casimir force only for $\sigma/\ell\lesssim 0.5$ and only for $a\lesssim 4\ell$. In this regime the unitarity argument based on transverse translational symmetry does not hold. In terms of the effective separation, the PFA to the roughness correction is approached from below with increasing correlation length. As pointed out at the end of Sect.~\ref{leadingcorr} it should be possible to intrinsically calibrate the absolute separation using  experimental results and take advantage of the much smaller corrections.
\begin{figure}
\includegraphics	[width=0.7\textwidth]{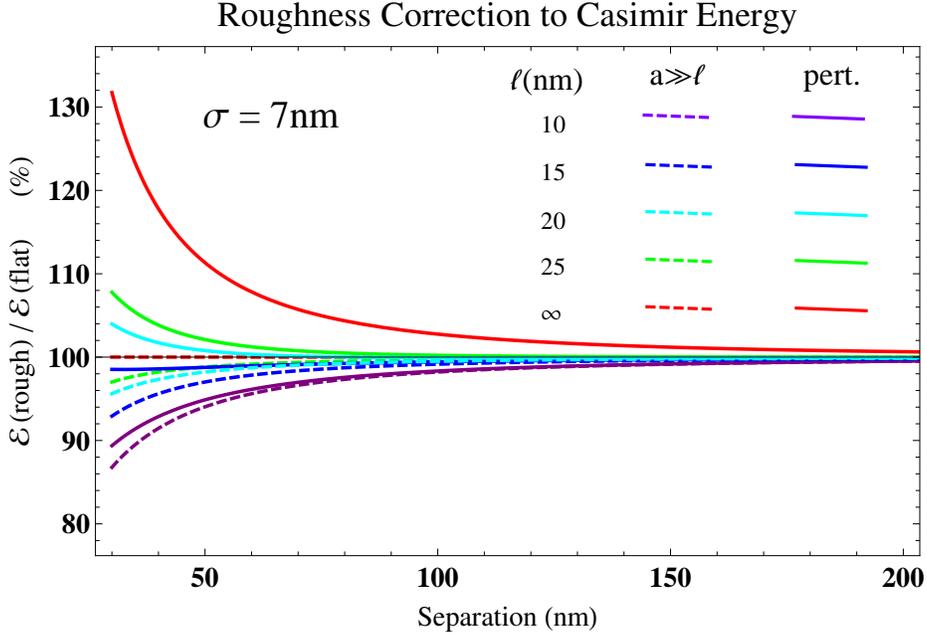}
\caption{(color online) Relative roughness corrections to the Casimir energy in \% due to a scalar satisfying Dirichlet boundary conditions on two plates, one of which is flat. The profile of the other is characterized by its variance $\sigma^2=49$nm$^2$ and correlation length $\ell$.  The separation is between equivalent planes representing the plates (see the text and \equ{aeff} for its relation to the mean separation.) The leading two-loop approximation for different correlation lengths $\ell$ is given by solid curves that correspond to those of fig.~\ref{PertCorrections}.  Dashed curves represent the correction in the effective low energy theory derived in the limit $a\gg\ell$. Pairs of dashed and solid curves of the same color correspond to the same correlation length $\ell=10\text{nm (violet)},15\text{nm (blue)}, 20\text{nm (cyan)}, 25\text{nm (green)  and } \ell=\infty$ (red). The leading two-loop approximation interpolates between the low-energy model for large separations $a\gg \ell$ and the PFA result (solid red) for small separations $a\lesssim \ell$. Note that typical roughness corrections are much smaller than the PFA suggests.}
\label{NonPert}
\end{figure}

We finally derived an effective low energy field theory that describes the limit $a\gg\ell$ with a single length parameter $\rho\sim\sigma^2/\ell$ characterizing the roughness of a plate. The correction in this limit indeed is described by an effective scattering matrix $t_\text{rough}$ for a plane displaced a distance $\rho/2$ from the mean of the profile as given in\equ{as}.
As illustrated by fig.~\ref{NonPert}, roughness in the effective low energy theory weakens the force at all separations and the effective scattering matrix is always less than for a flat plate of the same material, approaching the scattering matrix of the flat plate for wavelengths $\kappa\lesssim 1/\rho$.  It is also evident from  fig.~\ref{NonPert} that the corrected 2-loop estimate interpolates between the low energy effective model and the PFA, approaching the former for small and the latter for large correlation length $\ell$. For realistic correlation lengths and variances of the profile, the roughness correction at most is a few percent at $a=100$nm for a scalar field satisfying Dirichlet boundary conditions. It is even less for semi-transparent materials.

At separations where roughness corrections are important, temperature corrections are small and vice versa. The numerical results for small separations shown in figs.~\ref{PertCorrections}~\&~\ref{NonPert} therefore do not show temperature corrections. However, the formalism and most of the equations we derived include them. In the experimentally relevant electromagnetic case this could be of interest in the transition region  $500$nm-$2\mu$ where both corrections are small but comparable.                 

\begin{acknowledgments}
We thank K.V. Shajesh for useful comments and enlightening discussions and suggestions. M.S. thanks S. Fulling and K.A. Milton for being invited to  the QV-meeting in Norman, Oklahoma, where an early version of this paper was presented and discussed. This work was supported by the National Science Foundation with Grant No. PHY0902054 .
\end{acknowledgments}

\appendix
\section{Free Energy of a Massless Scalar Field for Two Flat Parallel Semitransparent Plates}
\label{AppA}
\subsection{An isolated flat semi-transparent plate}
Although this contribution to the free energy does not depend on the separation $a$ of two flat plates, it is finite and does depend on the temperature. We compute it for the sake of completeness.

Using Matsubara's formalism one\cite{Shajesh20111} readily finds that the irreducible contribution to the Helmholtz free energy per unit area, $f^{(1)}$, of a massless scalar field due to a semi-transparent flat plate of area $A$ described by the potential interaction $V(z)=\lambda \delta(z)$ is given by,
\bel{freeplate0}
f^{(1)}(T,\lambda)= \frac{T}{2}\sum_{n=-\infty}^\infty\int\frac{d\bk}{(2\pi)^2}\ln(1+\frac{\lambda}{2\kappa_n})\ ,
\ee
where $T$ is the temperature and $\kappa_n^2=(2\pi n T)^2+\bk^2$. Poisson's resummation formula allows one to rewrite\equ{freeplate0}
in the form,
\bal{freeplatePoisson}
f^{(1)}(T,\lambda)&=& \frac{1}{2}\sum_{n=-\infty}^\infty \int_{-\infty}^\infty\frac{d\zeta}{2\pi} \,e^{i n \zeta/T}\int\frac{d\bk}{(2\pi)^2}\ln(1+\frac{\lambda}{2\kappa})\nonumber\\
&=& \sum_{n=1}^\infty \frac{T}{2 \pi^2 n}\int_0^\infty d\kappa\,\kappa \sin(n\kappa/T) \ln(1+\frac{\lambda}{2\kappa})\\
&=& \frac{T^3}{2 \pi^2}\sum_{n=1}^\infty \frac{1}{n^3}\int_0^\infty dx\,x \sin(x) \ln(1+\frac{\lambda n}{2 T x})\nonumber\ ,
\ea
where the divergent, but temperature-independent, $n=0$ summand has been dropped by requiring that the free energy vanish at $T=0$. This amounts to ignoring the divergent change in zero-point energy due to insertion of a semi-transparent plate. In deriving the second expression of\equ{freeplatePoisson} we introduced spherical coordinates with $\kappa^2=\zeta^2+\bk^2$ and performed the angular integrations. The final expression in\equ{freeplatePoisson} is in fact finite. We may perform the summation and reduce the expression for the free energy per unit area of a flat plate to a single integral,
\bal{freeplate1}
f^{(1)}(T,\lambda)&=&\frac{T^3}{2 \pi^2}\int_0^\infty\frac{dy}{y}\left[\sum_{n=1}^\infty \frac{1-e^{-n y\lambda/(2T)}}{n^3}\right]\int_0^\infty dx\,x \sin(x) e^{-x y}\nonumber\\
&=&\frac{T^3}{\pi^2}\int_0^\infty\frac{dy}{(1+y^2)^2}\left[\zeta(3)-\text{Li}_3(e^{-y\lambda/(2T)})\right]>0\ .
\ea
The asymptotic behavior of $f^{(1)}$ is readily found,
\bal{AsympD}
f^{(1)}(T\ll\lambda)&\sim &\frac{T^3}{4\pi}\,\zeta(3)\\
\label{AsympHT}
f^{(1)}(\lambda\ll T)&\sim &\frac{T^2\lambda}{24}
\ea

For Dirichlet boundary conditions ($\lambda\rightarrow\infty$), the asymptotic expression in\equ{AsympD} holds at any temperature.\equ{AsympHT} is accurate to leading order in $\lambda$ for a weakly interacting plate. Note that the free energy of a single semi-transparent plate is positive and \emph{increases} monotonic with temperature for any value of $\lambda$. The corresponding contribution to the entropy therefore \emph{decreases} with increasing temperature. However, this ignores the bulk contribution to the entropy which generally overwhelms this reduction.  Including the bulk contribution, the total entropy due to insertion of  a Dirichlet plate is negative only for $ 1/T>\frac{(2\pi)^3}{135} V/A\sim 2 V/A$. It is negative only when the boundary of the container (on average) is within a thermal wavelength of the plate. Ignoring the finite size of the container in obtaining the entropy change due to the plate is no longer warranted in this situation. Although we here do not quantify the correction, it very likely is perfectly consistent that the entropy change due to insertion of a single plate is negative and decreases as the temperature increases. The negative contribution to the entropy can be qualitatively understood by the fact that the energy difference for excited cavity states increases upon insertion of the plate and the occupation numbers for excited states therefore decrease.

\subsection{Irreducible contribution to the free energy of a scalar due to two flat parallel semi-transparent plates}

We again use Matsubara's formalism and proceed as for a single plate. The irreducible contribution to the free energy per unit area, $f^{(2)}$, due to two semi-transparent parallel plates at separation $a$ is given by,
\bel{parallel0}
f^{(2)}(T;\lambda,\bar\lambda,a)= \frac{T}{2}\sum_{n=-\infty}^\infty\int\frac{d\bk}{(2\pi)^2}\ln(\Delta(\kappa_n))= \frac{T}{4\pi}\sum_{n=-\infty}^\infty\int_{2\pi |n| T} \kappa d\kappa\ln(\Delta(\kappa))\ ,
\ee
where $\kappa_n^2 = (2\pi n T)^2+\bk^2$ as before and $\Delta(\kappa)$ is given by\equ{Deltat}.  Contrary to the irreducible contribution from a single plate, $f^{(2)}$ is finite for any separation $a>0$. We again use Poisson's resummation formula to express the free energy in dual variables,

\bal{parallelPoisson}
f^{(2)}(T;\lambda,\bar\lambda,a)&=& \frac{1}{2}\sum_{n=-\infty}^\infty \int_{-\infty}^\infty\frac{d\zeta}{2\pi} \,e^{i n \zeta/T}\int\frac{d\bk}{(2\pi)^2}\ln(\Delta(\kappa))\nonumber\\
&&\hspace{-4em}=\frac{1}{2 \pi^2}\int_0^\infty d\kappa\,\kappa \left(\frac{\kappa}{2}+T\sum_{n=1}^\infty\frac{\sin(n\kappa/T)}{n}\right) \ln(\Delta(\kappa))\\
&&\hspace{-4em}=\frac{T}{2 \pi}\int_0^\infty d\kappa\,\kappa\, N(\frac{\kappa}{2\pi T})\, \ln\left(1-\frac{\lambda\bar\lambda e^{-2a\kappa}}{(\lambda+2\kappa)(\bar\lambda+2\kappa)}\right)\nonumber\ .
\ea
Here $N(x)$ is the staircase function ($[x]$ denoting the largest integer less than $x$),
\bel{staircase}
N(x):=1/2+[x]=x+\frac{1}{\pi}\arctan(\cot(\pi x))\ .
\ee
At low temperatures $f^{(2)}$ behaves as,
\bel{lowT}
f^{(2)}(2\pi T \tilde a\ll 1 ;\lambda,\bar\lambda)\sim \frac{1}{4\pi^2} \int_0^\infty d\kappa\,\kappa^2\ln(\Delta) + A\tilde a\frac{\pi^2 T^4}{90}\ ,
\ee
where the effective separation $\tilde a=a+\frac{1}{\lambda}+\frac{1}{\bar\lambda}$. The first term is just the Casimir energy of two semi-transparent plates \cite{Cavero20081,*Cavero20082}. Note that the $T^4$ behavior of the second term is the same as that of the bulk contribution to the free energy. In the Dirichlet limit $\lambda,\bar\lambda\sim \infty$ it simply subtracts the contribution to the free energy from the volume between the two plates. This again is qualitatively caused by the increased energy difference to excited states between the plates. The second term in\equ{lowT} is not correct in the weak coupling limit when $2\pi T\gg \lambda,\bar\lambda$. In the range $\lambda,\bar\lambda\ll 2 \pi T \ll 1/a$ we have that
\bal{lowTweak}
f^{(2)}(\lambda,\bar\lambda\ll 2 \pi T \ll 1/a)&\sim&\\
 &&\hspace{-12em}\sim\frac{\lambda\bar\lambda}{32\pi^2 a} \left(1+2\pi Ta\left(\frac{\bar\lambda\ln(T/\bar\lambda)-\lambda\ln( T/\lambda)}{\bar\lambda-\lambda}+1.27036\right)-\frac{19}{12}(2\pi T a)^2+\dots\right.\nonumber
\ea
Note that for weak coupling the entropy apparently diverges like $\ln(T)$ for small $T$. However, there is no violation of Nernst's theorem in this case, because\equ{lowTweak} only holds for $2\pi T\gg \lambda,\bar\lambda$. For lower temperatures\equ{lowT} is valid and the entropy vanishes proportional to $T^3$. The first term of\equ{lowTweak} reproduces the leading term of the Casimir energy for two weakly interacting parallel plates\cite{Cavero20081,*Cavero20082,Shajesh20111}.

The total free energy ${\cal F}^\parallel$, of a massless scalar field in the presence of two parallel flat plates is the sum of the bulk contribution, the irreducible one-body contributions of the individual plates in\equ{freeplate1} and the irreducible two-body contribution of\equ{parallelPoisson},
\bel{totalfree}
{\cal F}^\parallel(T;\lambda,\bar\lambda,a)=-V\frac{\pi^2 T^4}{90} + A f^\parallel(T;\lambda,\bar\lambda,a)\ ,
\ee
with
\bel{freeparallel}
f^\parallel (T;\lambda,\bar\lambda,a)=f^{(1)}(T,\lambda)+f^{(1)}(T,\bar\lambda)+ f^{(2)}(T;\lambda,\bar\lambda,a)\ .
\ee
We have chosen the normalization of the free energy so that ${\cal F}^\parallel$ vanishes at $T=0$ for widely separated plates, thus absorbing a divergent, but temperature and separation independent, factor in the normalization of the generating function.

\section{Thermal Green's Function of a Scalar in the Presence of Two Parallel Semitransparent Plates}
\label{AppB}
In Matsubara's formalism\cite{Fried1972bk,*Becher1984bk,*Kapusta1989bk} thermal Green\rq{}s functions of a mode at temperature $T$ are given by evaluating Euclidean Green's functions at the corresponding Matsubara frequency $\xi_n=2\pi n T$. We thus can draw on the literature for the Euclidean Green's function of a massless scalar in the presence of parallel semitransparent plates\cite{Cavero20081,*Cavero20082,Bordag2009bk,Shajesh20111}. The physical solution to\equ{gdiff} is,
\bal{g2plates}
g^\parallel(z,z';\kappa)&=& \frac{e^{-\kappa|z-z'|}}{2\kappa}-\frac{\Delta^{-1}}{2\kappa}[e^{-\kappa|z-a|},e^{-\kappa|z|}]\cdot
\left[ \begin{array}{cc}
t & -t e^{-\kappa a}\bar t\\
-t e^{-\kappa a}\bar t & \bar t
\end{array}\right]\cdot
\left[\begin{array}{c}
e^{-\kappa|z'-a|}\\
e^{-\kappa|z'|}
\end{array}\right]\\
&&\hspace{-6em}=\frac{e^{-\kappa|z-z'|}}{2\kappa}-\frac{\Delta^{-1}}{2\kappa}
\left(t e^{-\kappa(|z'-a|+|z-a|)}-t \bar t (e^{-\kappa(|z'|+a+|z-a|)}+e^{-\kappa(|z'-a|+a+|z|)})+\bar t e^{-\kappa(|z'|+|z|)}\right),\nonumber
\ea
\bel{Deltat}
\text{with   }\Delta(\kappa) =1-t\bar t e^{-2\kappa a}\ ,\ \ t=\frac{\lambda}{2\kappa+\lambda} \ \text{  and   }  \bar t=\frac{\bar\lambda}{2\kappa+\bar\lambda}\ .
\ee
Of particular interest to us is the correlation function in momentum space at $z=z'=a$ and its derivatives ($\phi^\prime_n(\bx,a)=\frac{\partial}{\partial a}\phi_n(\bx,a), \phi^{\prime\prime}_n(\bx,a)=\frac{\partial^2}{\partial a^2}\phi_n(\bx,a)$, etc.),
\begin{subequations}
\begin{align}
\int\hspace{-.4em} d\bx\, e^{-i\bk\bx}&\vev{\phi_n(\bx,a)\phi_n(0,a)}^\parallel&=&\lim_{z,z'\to a}g^\parallel(z,z';\kappa_n)=\frac{1}{\lambda}-\frac{2\kappa t}{\lambda^2\Delta}\Big|_{\kappa=\kappa_n} ,\label{valGa}\\
\int\hspace{-.4em} d\bx\, e^{-i\bk\bx}&\vev{\phi_n(\bx,a) \phi^\prime_n(0,a)}^\parallel&=&\lim_{z,z'\to a}\partial_{z'}g^\parallel(z,z';\kappa_n)=\frac{\kappa t \bar t e^{-2\kappa a}}{\lambda\Delta}\Big|_{\kappa=\kappa_n} ,\label{valpGa}\\
\int\hspace{-.4em} d\bx\, e^{-i\bk\bx}&\vev{\phi^\prime_n(\bx,a)\phi^\prime_n(0,a)}^\parallel&=&\lim_{z,z'\to a}\partial_z\partial_{z'}g^\parallel(z,z';\kappa_n)=\frac{\kappa^2}{\lambda}-\frac{\kappa}{2 t\Delta}\Big|_{\kappa=\kappa_n} ,\label{valpGpa}\\
\int\hspace{-.4em} d\bx\, e^{-i\bk\bx}&\vev{\phi_n(\bx,a)\phi^{\prime\prime}_n(0,a)}^\parallel&=&\lim_{z,z'\to a}\partial^2_{z'}g^\parallel(z,z';\kappa_n)=\frac{\kappa^2}{\lambda}-\frac{2\kappa^3 t}{\lambda^2\Delta}\Big|_{\kappa=\kappa_n} ,\label{valppGa}\\
\int\hspace{-.4em} d\bx\, e^{-i\bk\bx}&\vev{\phi^{(j)}_n(\bx,a)\phi^{(l)}_n(0,a)}^\parallel&=&\ \kappa_n^2\int\hspace{-.4em} d\bx\, e^{-i\bk\bx}\vev{\phi^{(j-2)}_n(\bx,a)\phi^{(l)}_n(0,a)}^\parallel,\label{valjp2Ga}
\end{align}
\label{valG}
\end{subequations}
where the expression is to be evaluated at the $n$-th Matsubara frequency ($\kappa\rightarrow\kappa_n=\sqrt{(2\pi n T)^2+\bk^2}$). The correlations in\equ{valG} are found by taking normal derivatives of\equ{g2plates} and using that $\lim_{s\to 0}\text{sign}(s)=0$, $\lim_{s\to 0}\text{sign}^2(s)=1$ and $\lim_{s\to 0}\delta(s)=\lim_{s\to 0}\text{sign}^\prime(s)=0$. \equ{valjp2Ga} expresses the fact that \equ{gdiff} relates correlations on the surface of the rough plate to ones with two fewer normal derivatives of $\phi$. Increasing the number of normal derivatives by two amounts to multiplying the Fourier-space correlation function by $\kappa^2$. The three correlation functions of Eqs.~(\ref{valGa}),~(\ref{valpGa})~and~(\ref{valpGpa}) thus generate all correlations with a higher number of normal derivatives such as\equ{valppGa}. This allows us to obtain Feynman rules for vertices with an arbitrary number of $h$-fields in Sect.~\ref{vertices}.

\bibliography{%
       biblio/b20100712-roughness%
              }

\end{document}